\documentclass[useAMS]{mn2e}
\usepackage{epsfig,times}
\def\epsfsize#1#2{\hsize}
\newcommand{\gapprox}{\,\rlap{\lower 2.5pt % > ungefaehr =
\hbox{$\sim$}}\raise 1.5pt\hbox{$>$}\,}
\newcommand{\lapprox}{\,\rlap{\lower 2.5pt % < ungefaehr =
\hbox{$\sim$}}\raise 1.5pt\hbox{$<$}\,}
\newcommand{\msun}{{M_{\odot}}}
\newcommand{\zsun}{{Z_{\odot}}}
\newcommand{\hsun}{\ensuremath{\rm H_{\odot}}}
\newcommand{\fesun}{\ensuremath{\textrm{Fe}_{\odot}}}

\newcommand{\hb}{{{\textrm{H}}\beta}}
\newcommand{\hda}{\ensuremath{{\textrm{H}}\delta_{A}}}
\newcommand{\hdf}{\ensuremath{{\textrm{H}}\delta_{F}}}
\newcommand{\hga}{\ensuremath{{\textrm{H}}\gamma_{A}}}
\newcommand{\hgf}{\ensuremath{{\textrm{H}}\gamma_{F}}}

\newcommand{\teff}{{T_{\rm eff}}}
\newcommand{\afe}{\ensuremath{[\alpha/{\rm Fe}]}}

\newcommand{\feh}{[{{\rm Fe}/{\rm H}}]}
\newcommand{\zh}{[{{\rm Z}/{\rm H}}]}

\newcommand{\lancon}{Lan\c{c}on}

\begin{document}

\title{\boldmath Evolutionary population synthesis: models, analysis of the ingredients and application to high-$z$ galaxies} 
\author[C. Maraston] {Claudia Maraston\thanks{Current address: University of Oxford, Denys Wilkinson Building, Keble Road, Oxford, OX1 3RH, UK}\\ 
Max-Planck-Institut f\"ur extraterrestrische Physik,
Giessenbachstra\ss e, D-85748 Garching b. M\"unchen, Germany\\}

\date {Accepted 2005 June 8. Received 2005 May 27; in original form 2004 October 5}
\pubyear{2005}

\maketitle

\begin{abstract}
Evolutionary population synthesis models for a wide range of
metallicities, ages, star formation histories, initial mass functions,
and Horizontal Branch morphologies, including blue morphologies at
high metallicity, are computed. The model output comprises spectral
energy distributions, colours, stellar M/L ratios, bolometric
corrections, and near-infrared spectral line indices. The energetics
of the post Main Sequence evolutionary phases are evaluated with the
fuel consumption theorem. The impact on the models of the stellar
evolutionary tracks (in particular with and without overshooting) is
assessed. We find modest differences in synthetic broad-band colours
as induced by the use of different tracks in our code (e.g., $\Delta
(V-K)\sim 0.08~\rm mag$; $\Delta(B-V)\sim 0.03~\rm mag$). Noticeably,
these differences are substantially smaller than the scatter among
other models in the literature, even when the latter adopt the same
evolutionary tracks. The models are calibrated with globular cluster
data from the Milky Way for old ages, and the Magellanic Clouds plus
the merger remnant galaxy NGC~7252, both for young ages of $\sim
0.1-2\;$Gyr, in a large wavelength range from the $U$-band to the
$K$-band. Particular emphasis is put on the contribution from the
Thermally-Pulsing Asymptotic Giant Branch phase. We show that this
evolutionary phase is crucial for the modelling of young stellar
populations by the direct comparison with observed spectral energy
distributions of Magellanic Clouds clusters, which are characterised
by relatively high fluxes both blueward and redward the $V$-band.  We
find that the combination of the near-IR spectral indices $\rm C_{2}$
and $\rm H_{2}O$ can be used to determine the metallicity of $\sim
1~\rm Gyr$ stellar populations.  As an illustrative application, we
re-analyze the spectral energy distributions of some of the high-$z$
galaxies ($2.4 \lapprox z \lapprox 2.9$) observed with the Spitzer
Space Telescope by Yan et al.~(2004). Their high rest-frame near-IR
fluxes are reproduced very well with the models including
Thermally-Pulsing Asymptotic Giant Branch stars for ages in the range
$\sim 0.6-1.5~\rm {Gyr}$, suggesting formation redshifts for these
objects around $z\sim 3-6$.
\end{abstract}
\section{Introduction}
The evolutionary population synthesis (EPS) is the technique to model
the spectrophotometric properties of stellar populations, that uses
the knowledge of stellar evolution. This approach was pioneered by
B.~Tinsley in a series of fundamental papers, that provide the basic
concepts still used in present-day computations. The models are used
to determine ages, element abundances, stellar masses, stellar mass
functions, etc., of those stellar populations that are not resolvable
in single stars, like galaxies and extra-galactic globular
clusters. Due to the ubiquitous astrophysical applications of EPS
models, a rich literature has been developed so far (Bruzual~1983;
Renzini \& Buzzoni~1986; Chiosi, Bertelli and Bressan~1988;
Buzzoni~1989; Charlot \& Bruzual~1991; Bruzual \& Charlot~1993;
Worthey~1994; Vazdekis et al.~1996; Tantalo et al.~1996; Fioc \&
Rocca-Volmerange~1997; Bressan, Granato \& Silva~1998; Maraston~1998;
Leitherer et al.~1999; Brocato et al.~2000; Thomas, Maraston \&
Bender~2003; Thomas, Maraston \& Korn~2004).

In the simplest flavour of an EPS model, called {\it Simple Stellar
Population} (hereafter SSP) by Renzini~(1981), it is assumed that all
stars are coeval and share the same chemical composition. The
advantage of dealing with SSPs is twofold.  First, SSPs can be
compared directly with globular cluster (GC) data, since these are the
``simplest'' stellar populations in nature. This offers the advantage
of {\it calibrating} the SSPs with those GCs for which ages and
element abundances are independently known, an approach introduced in
the review by Renzini \& Fusi Pecci (1988). This step is crucial to
fix the parameters that are used to describe that part of the model
``input physics'' - convection, mass loss, mixing - that cannot be
derived from first principles.  The calibrated models can be applied
with more confidence to the study of extragalactic stellar
populations. This step is taken in the models of Maraston~(1998) and
in their extension presented here. Second, complex stellar systems
which are made up by various stellar generations are modelled by
convolving SSPs with the adopted star formation history
(e.g. Tinsley~1972; Arimoto \& Yoshii 1986; Rocca-Volmerange \&
Guiderdoni 1987; Vazdekis et al.~1996; Kodama \& Arimoto~1997; Barbaro
\& Poggianti 1997; Bruzual \& Charlot~2003). Therefore the deep
knowledge of the building blocks of complex models is very important.

Two techniques are adopted to compute SSP models, which differ
according to the integration variable adopted in the post-Main
Sequence: isochrone synthesis and `fuel consumption based'
algorithms. With the `isochrone synthesis' (e.g. Chiosi, Bertelli \&
Bressan~1988; Charlot \& Bruzual~1991) the properties of a stellar
population are calculated by integrating the contributions to the flux
in the various passbands of all mass-bins along one isochrone, after
assuming an Initial Mass Function (IMF). Usually isochrones are
computed up to the end of the Early Asymptotic Giant Branch
phase. Later stellar phases like the Thermally-Pulsing Asymptotic
Giant Branch are added following individual recipes or are neglected.

In the `fuel consumption' approach (Renzini 1981; Renzini \& Buzzoni
1986; Buzzoni~1989; Maraston~1998), the integration variable in Post
Main Sequence is the so-called {\it fuel}, that is the amount of
hydrogen and/or helium that is consumed via nuclear burning during a
given post Main Sequence phase. The fuel at a given age is computed on
the stellar evolutionary track of the {\it turnoff} mass (i.e., the
mass completing the hydrogen-burning phase), thereby neglecting the
dispersion of stellar masses in post main sequence. However since a
mass difference of only few percent exists between the {\it turnoff}
mass and the mass at any other post-Main Sequence phase, this
assumption can be made safely, as also shown by Charlot \&
Bruzual~(1991). The advantages of the fuel consumption approach are of
two kinds. First, the fuel as integration variable is very stable
since it is directly proportional to the contributions of the various
phases to the total luminosity. This is very important in luminous,
but short-lived evolutionary stages, e.g. the bright Red Giant Branch
phase, where the evolutionary mass practically does not change. We
note that the problem of the numerical instability on the RGB was
early recognized by Tinsley \&~Gunn (1976). Second, and more
important, there are several relevant stellar phases (e.g. blue
Horizontal Branch, Thermally Pulsing Asymptotic Giant Branch, very hot
old stars etc.)  whose theoretical modeling is uncertain because of
mass loss and for which complete stellar tracks are not available. The
fuel consumption provides useful analytical relations that link the
Main Sequence to the post Main Sequence evolution, by means of which
one can include into the synthesis the energy contributions of these
uncertain phases using, e.g. observations. The `isochrone synthesis'
technique is used in all models in the literature, with the exception
of the models by Buzzoni~(1989), Maraston~(1998) and those presented
here, that adopt the fuel consumption theorem.

Besides the method used to compute them, evolutionary population
synthesis models keep the uncertainties inherent in the stellar
evolutionary tracks and in the spectral transformations. Charlot,
Worthey \& Bressan~(1996) investigated both classes of uncertainties
by comparing their EPS models. The analysis is very illustrative for
the three considered EPS, however does not yield information about the
sole impact of the stellar tracks. Here we make a different exercise,
because we can use the same EPS code and just vary the model
ingredients. In this way we can isolate their impact on the final
model. To quantify the model uncertainties is indeed very important
since the cosmological inferences that are derived on the basis of
galaxy ages and metallicities rely ultimately on the stellar
population models.

Maraston (1998) presents a fuel-consumption-based code for
evolutionary population synthesis, and SSP models for solar
metallicity, and ages from 30 Myr to 15 Gyr. Distinct features of that
work are:
\begin{enumerate}
\item the extension of the fuel consumption theorem to
compute models for young and intermediate-range ages; 
\item the inclusion of a well-calibrated semi-empirical
Thermally-pulsing Asymptotic Giant Branch phase, and the computation
of realistic colors of intermediate-age ($t\lapprox~2~\rm Gyr$)
stellar populations
\item the modular structure of the code, that allows experiments with
the EPS ingredients.
\end{enumerate}
Subsequentely the code has been updated to the computations of SSP
models covering a wide range in metallicities and the model output
have been extended to e.g. the spectral energy distributions, spectral
indices, redshift evolution, with several applications being already
published (Maraston \& Thomas 2000; Saglia et al.~2000; Maraston et
al. 2001; Maraston et al.~2003; Thomas, Maraston \& Bender ~2003;
Thomas, Maraston \& Korn 2004; Ferraro et al.~2004). This work is
devoted to discuss comprehensively the overall EPS models and
ingredients, in particular, the inclusion of the TP-AGB phase in the
synthetic spectral energy distributions.

The paper is organized as follows. In Section~2 the properties of the
evolutionary population synthesis code are recalled. The model
ingredients, i.e. the fuel consumptions, the temperature distributions
and the model atmospheres, for the various metallicities and ages, are
described in detail in Section~3. In particular, the recipes for
Horizontal Branch and Thermally Pulsing Aymptotic Giant Branch are
presented in Section 3.4. The various model output are discussed in
Section~4, where the comparisons with observational data and with
models from the literatures are also presented. Section~5 deals with
the model uncertainties, while Section~6 presents an high-redshift
application of the model SEDs in which the TP-AGB has a primary
importance. Finally a summary of the main results is given in
Section~7.
\section{Algorithm}
In this section we recall the basic equations that define the
algorithm of the EPS code by Maraston (1998, hereafter M98).
Following Renzini \& Buzzoni~(1986, hereafter RB86), the total
bolometric luminosity of a SSP of age $t$ and chemical composition
$[Y,Z]$, is splitted conveniently into the contributions by Main
Sequence (MS) and post Main Sequence (PMS)
\begin{equation}
 L_{\rm SSP}^{\rm bol}(t; [Y,Z]) = L_{\rm MS}^{\rm bol}(t; [Y,Z]) + L_{\rm PMS}^{\rm bol}(t; [Y,Z]).
\end{equation}
\label{lt}
because the two quantities depend on different ingredients. The MS
light is produced by stars spanning a very wide mass range, while the
PMS one is produced by stars of virtually the same mass. Therefore $
L_{\rm MS}^{\rm bol} $ depends on the adopted mass-luminosity relation
$L(M,t)$ and IMF, $\Psi(M)=A{M}^{-s}$ ($A$ being the scale factor
related to the size of the stellar population), while $ L_{\rm
PMS}^{\rm bol}$ depends on the nuclear fuel burned in the various
evolved stellar stages. The MS luminosities are computed by
integrating the contributions by each mass bins along a isochrone,
from a lower mass limit ${M_{\rm inf}}$ (usually $0.1~\msun$) to the
current turnoff mass $M_{\rm TO}$(t), having assumed an IMF $\Psi(\rm
M)$
\begin{equation}
L_{\rm MS}^{\rm bol}(t; [Y,Z])=\int_{\rm M_{inf}}^{\rm M_{TO}}L({\rm
M}(t; [Y,Z]))\psi({\rm M})dM
\label{lms}
\end{equation}
The PMS luminosity contributions are computed by means of the {\it
Fuel Consumption Theorem} (RB86)
\begin{equation}
L_{\rm PMS}^{\rm bol}(t; [Y,Z])=9.75\cdot 10^{10}\cdot b(t)\cdot
\sum_{\rm J} {\rm Fuel}_{\rm J}(M_{\rm TO}(t; [Y,Z]))
\label{lpms}
\end{equation}
where the {\it evolutionary flux} $b(t; [Y,Z])$,
\begin{equation}
b(t; [Y,Z])=\psi(M_{\rm TO}(t; [Y,Z]))|{\dot{M}}_{\rm TO}|
\label{biditi}
\end{equation}

provides the rate of stars evolving to any PMS phase $\rm j$ at the
age $t$ of the stellar population, and ${\rm Fuel}_{\rm J}(M_{\rm
TO}(t; Y,Z))$ is the amount of stellar mass to be converted in
luminosity in each of these phases. The multiplicative factor in
Eq.~\ref{lpms} stems from expressing the luminosity in solar units,
the evolutionary flux $b(t)$ in years and the fuel in solar masses
through the Einstein equation $E=\Delta M \cdot c^2$, with
$\Delta=6.55\cdot10^{-3}$. The latter is derived by considering that
the transformation of 1 g of H into He releases $\sim 5.9\cdot
10^{18}$ erg. This average value takes into account the dependence on
whether the CNO cycle or the pp chain is at work and the different
neutrino losses as a function of the temperature of the burning
(Renzini 1981).
\section{Ingredients}
\label{ingredients}
Following M98, the ingredients of evolutionary synthesis models are:
\begin{enumerate} 
  \item {\it The energetics}: mass-luminosity relations for the MS and
  fuel consumptions for PMS phases;
\item {\it The surface parameters}: the effective temperatures and
surface gravities of the evolutionary phases;
\item {\it The transformations to observables}: spectra, or colours
and bolometric corrections as functions of gravity and temperature, to
convert the bolometric luminosity into a spectral energy distribution.
\end{enumerate}
The key feature of the code is to have the three ingredients being
allocated in three {\it independent} sets of matrices. This is very
convenient as the code can be used to understand the impact of the
various input {\it selectively} on the final result. We will use this
structure to understand the discrepancies between EPS models that are
based on different stellar evolutionary tracks.
The adopted ingredients are described in the next subsections.
\subsection{Energetics}
\label{energetics}
The first matrix contains the energetics, i.e. the luminosities of MS
stars and the fuel consumptions of PMS phases. In general, both are
taken from stellar evolutionary models, except for those stellar
phases that are poorly understood (Section~\ref{recipes}), for which
the energetics are estimated semi-empirically, by means of
observations and with the aid of the fuel consumption theorem.
\subsubsection{Stellar models}
\label{tracks}
The bulk of input stellar models (tracks and isochrones) is from
Cassisi et al.~(1997a,b; 2000; see also Bono et al.~1997). Their main
features are summarized in the following. These are {\it canonical}
stellar evolutionary tracks, i.e. the efficiency of the overshooting
parameter is assumed to be zero. The actual size of the overshooting
is a matter of debate since several years and work is in progress to
calibrate the overshooting parameter with observational data (Bertelli
et al.~2003; Woo et al.~2003). These articles favour moderate amounts
of overshooting, but the results on different stellar evolutionary
tracks are discrepant. The Cassisi tracks are used to compute what we
will refer to as {\it standard} SSP models. The choice of these tracks
as basis is due to the following reasons: i) extensive calibrations
with galactic globular clusters (GCs) have been performed (Cassisi \&
Salaris 1997; Cassisi, Degl'Innocenti, Salaris 1997; De Santi \&
Cassisi 1999); ii) tracks (isochrones) are provided with very fine
time (mass) spacing (e.g., a typical Red Giant Branch track contains
$\sim 5000$ models) which is essential to perform good numerical
integrations; iii) these tracks are the closest to the ones (from
Castellani, Chieffi \& Straniero~1992) that were adopted for the solar
metallicity models presented in M98. For sake of homogeinity, the M98
models have been re-computed with the solar metallicity tracks of the
Cassisi's database. Minimal differences have been found, that are due
to the temperature/colour transformations rather than to the stellar
tracks. The metallicity of the Cassisi tracks range from $\zsun/200$,
typical of the Milky Way halo to $2\zsun$, the helium enrichment law
being ${\Delta Y/\Delta Z}\sim 2.5$. In order to extend the
metallicity range, we implement a set of tracks with 3.5 solar
metallicity, and the same ${\Delta Y/\Delta Z}$, from the Padova
database (see below). The exact values of helium and metals for the
SSP grid are tabulated in the section presenting the results.

Most SSP models in the literature are based on the tracks by the
Padova group (Fagotto et al.~1994; Girardi et al.~2000; Salasnich et
al.~2000). Therefore it is interesting to explore the effects of other
stellar evolutionary tracks, on the ages and metallicities inferred
for real stellar populations. To this aim several SSPs have been
computed by means of the Padova stellar models. The various
comparisons will be shown in Section~\ref{uncertainties}. The issue is
a very important one as at metallicities above solar, i.e. in the
range more relevant to massive galaxies, the calibration of the tracks
is hampered by the lack of GCs with ages and chemical compositions
known independently. 

Finally the solar metallicity isochrones/tracks of the Geneva database
(Schaller et al.~1992; Meynet et al.~1994) are adopted in order to
compute very young SSPs ($10^{-3}\leq t/{\rm Myr} < 30$).
\subsubsection{Main Sequence: Mass Luminosity relations}
\label{ms}
Isochrones are adopted up to the turnoff, and the MS luminosity
contributions are evaluated by means of Equation~\ref{lms}. Therefore
the results depend on the mass-luminosity relations of the isochrones.
\begin{figure}
 \psfig{figure=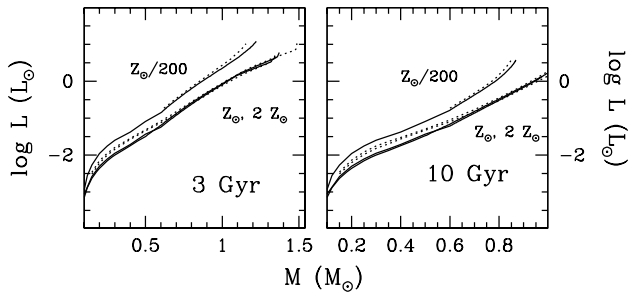,width=\linewidth}
  \caption{The relation between luminosity and mass for the isochrones
  of Cassisi (solid lines) and Padova (dotted lines). The
  metallicities are labelled. Note that the relations for solar and
  twice solar metallicities nearly overlap.}
\label{mlrel}
\end{figure}
Figure~\ref{mlrel} compares the mass-luminosity relations of the
isochrones from Cassisi and Padova (solid and dotted lines,
respectively) for various ages and metallicities. In general, a fairly
good agreement is found for masses $\gapprox~0.5\msun$, independent of
the metallicity, while the low MS of the Padova tracks with high
metallicity (solar and above) is brighter than that of the Cassisi
tracks, by nearly a factor 2. However this effect is not important,
because the contribution of the low MS to the total light is very
small, unless the stellar population has a very steep IMF
(e.g. $s\gapprox 3.5$ in the notation in which the Salpeter exponent
is 2.35, M98), so that its light is dwarf dominated. Noticeable is
instead the effect of overshooting, because of which the Padova MS has
a turnoff mass at given age that is {\it larger} than that of
canonical tracks (e.g. at 3 Gyr and solar metallicity the turnoff
masses are $1.45~\msun$ and $1.37~\msun$, respectively). Stellar
models with overshooting have more massive convective cores, therefore
they run to higher luminosities and live longer than classical
models. This effect stems from the higher fuel for given mass when
overshooting is considered, which prolongues the MS lifetime. The
effect lasts until the mass has a convective core on the MS, i.e. it
disappears for $M\lapprox~1~\msun$.

As to metallicity effects, at given mass a higher metal content makes
the star fainter, because of the combined effects of the less amount
of hydrogen and the higher opacity.  For example, a $0.8~\msun$ star
with~ $2~\zsun$ metallicity is a factor 3 fainter on the MS than one
with the same mass but metallicity~$\zsun/200$. Instead, there is no
difference between the solar and twice-solar metallicity MS
relations. This comes from the fact that helium increases along with
metallicity in both tracks, according to the helium enrichment law
$\Delta Y/\Delta Z\sim 2.5$. The higher helium at larger metallicity
counterbalances the metallicity effects and keeps the star at roughly
the same brightness.
\begin{figure}
 \psfig{figure=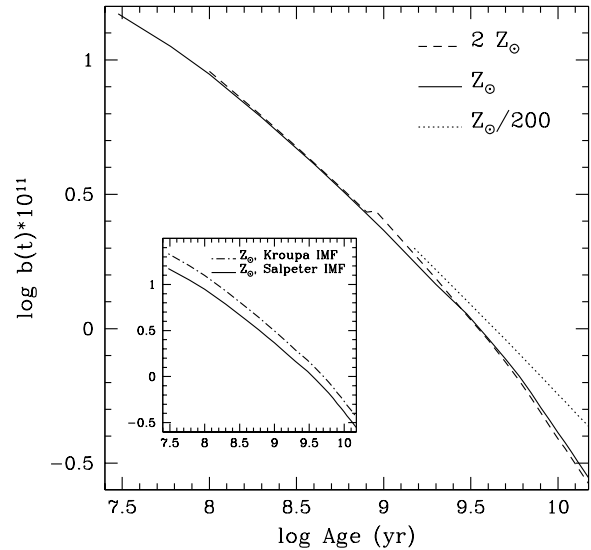,width=\linewidth}
  \caption{The evolutionary flux $b(t)$ (Eq.~\ref{biditi}) that
  measures the rate of evolution off the MS, for various
  metallicities. The $b(t)$ are normalized by means of the IMF scale
  factor $A$ in such a way as to be referred to 1 $\msun$. The IMF
  dependence is highlighted in the small internal box, where the
  $b(t)$ for solar metallicity are shown for the Kroupa and the
  Salpeter IMF. The Kroupa SSP has a $\sim$ 30 per cent higher
  evolutionary rate with respect to the Salpeter SSP.}
\label{bflux}
\end{figure}

In order to link the PMS evolution of the turnoff mass to its MS, one
needs to know the rate of evolution of turnoff-like stars off the MS
through the later evolutionary stages, as a function of the SSP
parameters (age, metallicity, IMF). According to RB86 this quantity is
expressed analytically by the evolutionary flux $b(t)$
(Eq.~\ref{biditi}), that is proportional to the time derivative of the
relation turnoff mass/age and the adopted IMF. The dependence of the
function $b(t)$ on the SSP parameters is shown in
Figure~\ref{bflux}. The main panel focuses on age and chemical
composition effects, the small one on the IMF. Obviously $b(t)$
depends mainly on the age of the SSP, as the derivative of the turnoff
mass does, while metallicity effects have a much milder influence. The
effect of the IMF on $b(t)$ was discussed in M98. In brief, the
flatter the IMF, the slower $\Psi(M)$ increases with time, so that the
rapid decrease in \.M$_{\rm TO}$ dominates for flat IMFs.  This
results in a steeper SSP luminosity evolution for flatter IMFs. In the
small box in Figure~\ref{bflux} we show $b(t)$ for a Kroupa~(2001)
IMF, that will be used as alternative choice to the Salpeter one to
compute SSP models. The Kroupa~(2001) IMF is described as a
multiple-part power law, with exponents: 1.3 for $0.1 \lapprox \rm
M/\msun \lapprox 0.5 \msun$, 2.3 for larger masses (in the notation in
which the Salpeter exponent is 2.35). The Kroupa IMF has a relatively
higher fraction of massive stars, which explains why the correspondent
$b(t)$ is steeper than that referred to the Salpeter IMF (note that
the $b(t)$'s are normalized to $1~\msun$ of total mass of the parent
SSP and it is the normalization factor that produces the difference).
\subsubsection{Post Main Sequence: Fuel comsumptions}
\label{fuels}
The luminosity contributions of the PMS phases are computed by means
of the fuel consumption theorem, according to Eq.~\ref{lpms}. The
nomenclature of the main PMS evolutionary phases, in order after the
MS, are: Sub Giant Branch (SGB); Red Giant Branch (RGB); Helium
Burning or Horizontal Branch (HB); Early Asymptotic Giant Branch
(E-AGB); Thermally Pulsing Asymptotic Giant Branch (TP-AGB).
\begin{figure}
 \psfig{figure=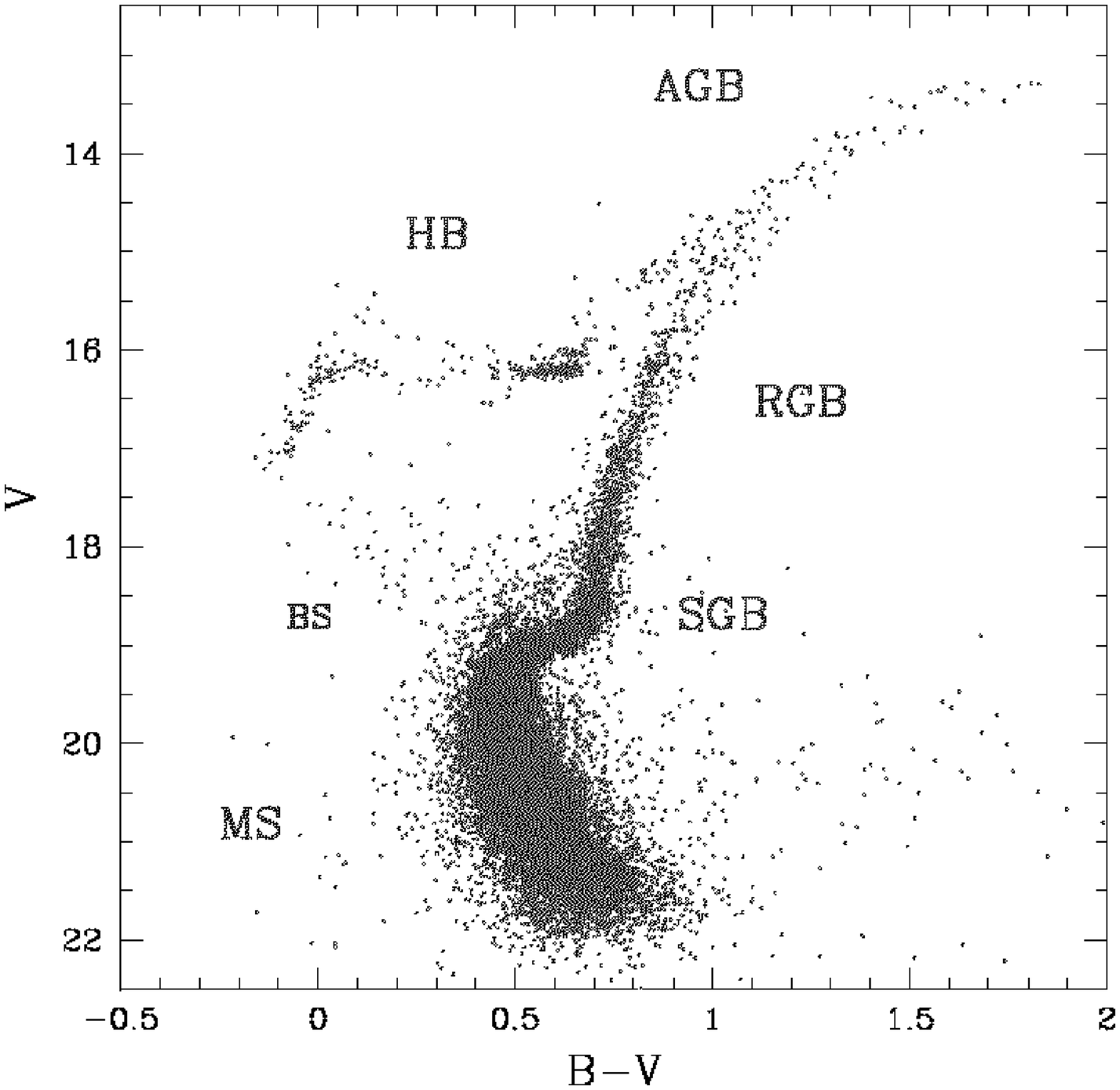,width=\linewidth}
  \caption{The stellar evolutionary phases indicated on the observed
  colour magnitude diagram of the metal-poor Milky Way globular
  clusters NGC~1851 (data from Piotto et al.~2002). BS stays for 'blue
  stragglers', a class of stars not included in the evolutionary
  synthesis. No separation is made between the early and the
  thermally-pulsing AGB.}
\label{ngc1851}
\end{figure}
\begin{figure*}
 \psfig{figure=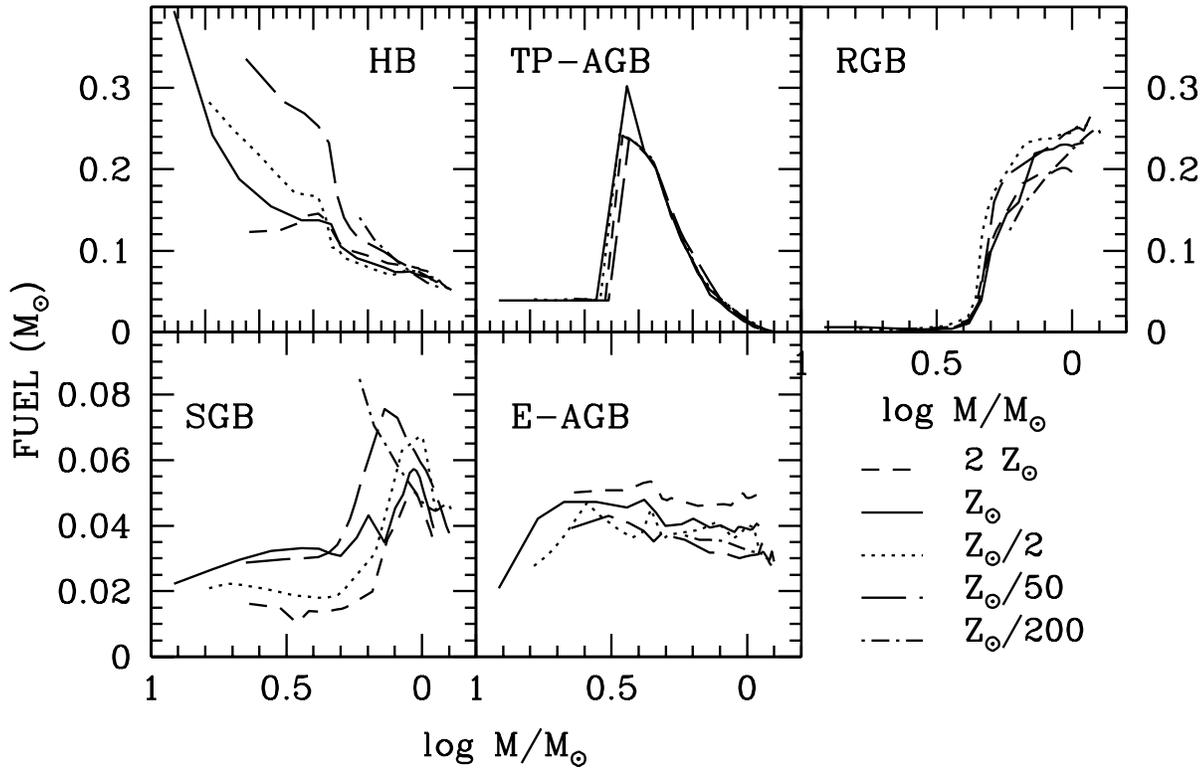,width=\linewidth}
  \caption{The fuel consumptions (in {\rm $M_{\odot}$}) in the various
  PMS phases, as functions of the turnoff mass, for various
  metallicities.}
\label{fuelfasi}
\end{figure*}
The stellar evolutionary phases are highlighted on the observed colour
magnitude diagram of NGC~1851, a metal-poor GC of the Milky Way (data
from Piotto et al.~2002).

In order to compute the fuels, evolutionary tracks are adopted up to
the completion of the E-AGB. Complete tracks for the TP-AGB phase are
not available. In fact it is difficult to follow the physics of the
stellar interiors during the thermal pulses and stellar tracks are
actually restricted to envelope models (e.g. Renzini \& Voli~1981;
Iben \& Renzini~1983; Lattanzio~1986; Boothroyd \& Sackmann~1988;
Bloecker \& Schoenberner~1991; Marigo, Bressan \& Chiosi~1996;
Wagenhuber \& Gronewegen~1998; Marigo 2001; Mouhcine \& Lan\c
{c}on~2002), and even the latter are uncertain due to the occurrence
of strong mass-loss that aborts the phase (the {\it superwind} phase,
Iben \& Renzini~1983). In the present models the energetics of the
TP-AGB comes from the semi-empirical calibration of M98 (see
Section~\ref{TP-AGB}).

The fuel for the evolutionary phase $\rm j$ (until the E-AGB) is
computed by integrating the product of the evolutionary time and the
emergent luminosity along the track appropriate to that phase. The
tracks for the given turnoff masses are obtained from interpolation in
log mass. The evolutionary mass for the helium burning phase $M_{\rm
HB}$ is obtained from $M_{\rm TO}$ after evaluation of the mass loss
during the RGB. This allows to play with various HB morphologies (see
Section~\ref{recipes}). Finally, the separation between HB and E-AGB
is set when the mass of the CO core along the track is different from
zero. Note that each phase is divided suitably into a certain number
of subphases, in order to map appropriately the spectral type changes
(M98). The criterion for the subdivision into suphases depends on the
temperature, therefore is described in Section~\ref{temperatures}.

Figure~\ref{fuelfasi} shows the fuels (in {\rm $M_{\odot}$}) for the
PMS evolutionary phases, as functions of the turnoff mass, for various
metallicities. As already pointed out by RB86, the most relevant PMS
phase changes with the evolutionary mass, i.e. with the age of the
stellar population, and we find here that the trend does not depend on
the chemical composition.
%age effects

In massive stars ($M\gapprox~3~\msun$), i.e. those dominating {\it
young} - $t\lapprox 0.2~{\rm Gyr}$ - stellar populations, the dominant
PMS phase is the HB, its fuel decreases with the decreasing stellar
mass. In low-mass stars ($M\lapprox~2~\msun$), i.e. those dominating
{\it old} - $t\gapprox 2~{\rm Gyr}$ - stellar populations, the RGB
phase is the most important PMS phase, when a He-degenerate core is
developed (RB86; Sweigart, Greggio \& Renzini~1989). Stars with masses
in the narrow mass range between 3 and 2~$\msun$, i.e. those
dominating $0.2 \lapprox t/{\rm Gyr} \lapprox 2$ old stellar
populations, spend a conspicuos amount of fuel on the TP-AGB phase.
The onset of the development of the TP-AGB and RGB phases has been
called by RB86 ``phase transitions''.
% metallicity effects

The HB fuel (left-hand upper panel) of massive stars is affected by
metallicity in the sense of a higher fuel at a lower metal content,
owing to the higher relative abundance of hydrogen, an effect similar
to what pointed out for the MS luminosity (Section~\ref{ms}). For
example, at masses $\gapprox~4 \msun$, the $\zsun/20$ metallicity has
nearly a factor 2 more fuel than the solar one. In the small mass
regime ($M \lapprox 1.5 \msun$, $t_{\rm SSP} \gapprox 3~\rm Gyr$)
metallicity effects are negligible.

\begin{figure}
 \psfig{figure=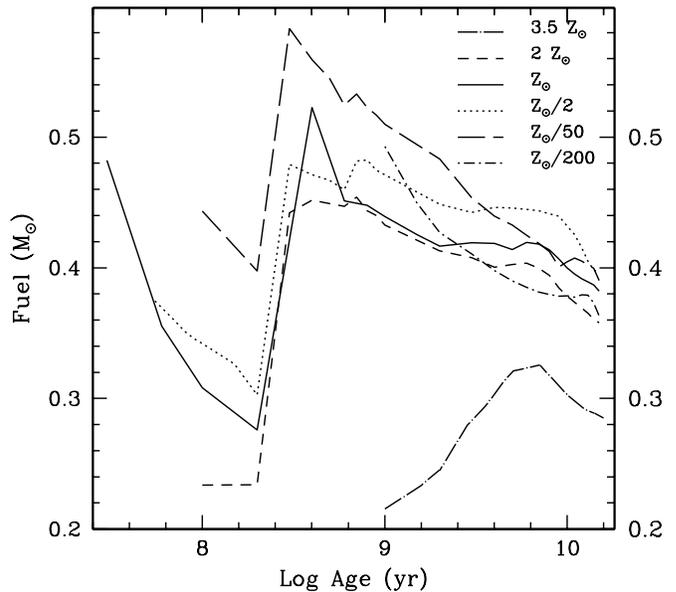,width=\linewidth}
  \caption{The total fuel consumptions (in {\rm $M_{\odot}$}) in the
  whole PMS, as a function of the age of the SSP, for the various
  metallicities. Note that the fuel scales directly with
  luminosity. The very-metal rich SSP of the Padova database
  (long-dashed/dotted line) sticks out for having a very small amount
  of fuel. The TP-AGB phase is not included in this SSP.}
\label{ftotage}
\end{figure}

The RGB phase (right-hand upper panel) starts to develop at masses
around $2~\rm \msun$ almost independent of metallicity in these
classical (no overshooting) tracks, in excellent agreement with the
early findings by Sweigart, Greggio \& Renzini~(1990).  The RGB fuel
is rather insensitive to metallicity until $\zsun$. For reference, a
10 Gyr, $ \zsun/200$ stellar population has $0.24~\msun$ of RGB fuel
and a coeval one with solar metallicity $0.23~\msun$. However, at
higher metallicities the RGB fuel starts decreasing with increasing
metal content, and for example a stellar population with $3.5~\zsun$
(see Figure~\ref{ftotage}) has $0.15~\msun$ of RGB fuel, nearly 35~per
cent less than the solar chemical composition. This is the effect of
the very high helium abundance associated to the high metal content
because of the helium enrichment law $\Delta Y/\Delta Z\sim 2.5$, that
reduces drastically the amount of hydrogen, i.e. the RGB source of
fuel. A discussion on the effect of helium abundance on the fuel of
PMS phases is found in Greggio \& Renzini~(1990).

The TP-AGB fuel (middle upper panel) is one of the most conspicuous,
together with that in HB and RGB, and is a strong function of the
stellar mass, therefore of the age of the stellar population. It
reaches a maximum for masses between 3 and 2 $\msun$ and is negligible
for masses outside this narrow intermediate-mass range. Furthermore,
the TP-AGB fuel does not depend appreciably on metallicity (see
Section~\ref{TP-AGB}).

Finally the SGB and E-AGB (lower panels) are the least important
phases, providing at most $\sim~20$ per cent and $\sim~10$ per cent,
respectively of the total PMS luminosity (see also
Section~\ref{resbolometric}).

Figure~\ref{ftotage} shows the total fuel consumption in the whole
PMS, as a function of age, for the various metallicities. In
interpreting this figure in a stellar population perspective it is
useful to remind that the fuel scales directly with the PMS luminosity
of a stellar population.

The effect of the chemical composition that we discussed previously is
evident on the total fuel of stellar populations with ages smaller
than $\sim~1~{\rm Gyr}$, that is larger the lower the metal content
is. A young metal-poor stellar population with $Z= \zsun /20$ has
$\sim$ 20 per cent more fuel to be burned, e.g. is 20 per cent
brighter than a coeval one in which the metallicity is twice solar. In
older stellar populations metallicity effects become less important,
as a consequence of the small metallicity dependence of the RGB fuel,
that is the largest source of energy at high ages. An exception is
however the very metal-rich stellar population ($Z=3.5~\zsun$,
dashed-dotted line), that at old ages has $\sim~25$ per cent less fuel
than all other chemical compositions, for which the fuel scatter
around a value $\sim~0.4~\msun$. This is due to the very high
abundance of helium in these tracks, caused by the assumed helium
enrichment law (${\Delta Y/\Delta Z}\sim 2.5$), that implies the
abundance of hydrogen to be only $0.45$. This explains the sharp
decrease in fuel since hydrogen is its most important source.

The sizable increase of PMS fuel at ages around 0.3 Gyr
($log~t\sim~8.47$) marks the onset of the AGB phase transition, that is
dominated by the TP-AGB (M98, Figure~\ref{fuelfasi}). At later ages
the RGB phase transition occurs, which is barely visible in
Figure~\ref{ftotage} as a small bump around 1 Gyr, due to the
simulatenous decrease of TP-AGB fuel. Finally, the bumps in the fuels
at late ages ($t\gapprox 6~{\rm Gyr}$) reflect the trend of the SGB
fuel (Figure~\ref{fuelfasi}).
\begin{figure}
 \psfig{figure=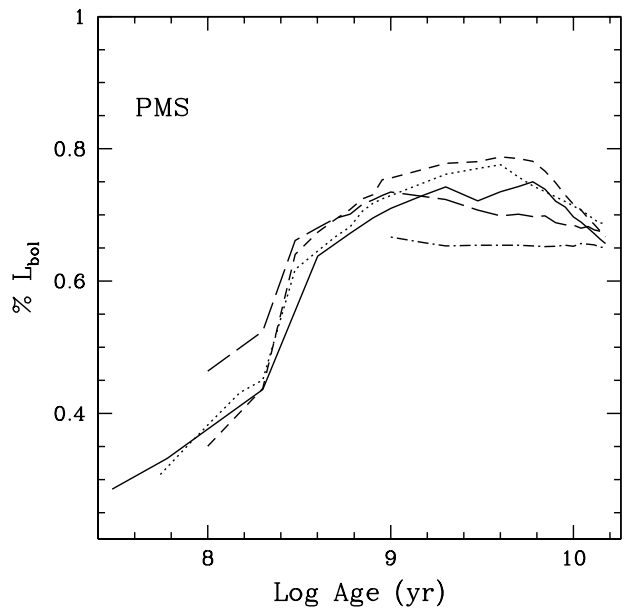,width=\linewidth}
  \caption{The luminosity contributions of the PMS (evaluated with
  Eq.~3) for the various metallicities (linestyles as in the previous
  plots) and Salpeter IMF.}
\label{bolmspms}
\end{figure}
After the development of the TP-AGB, $\sim~70$ per cent of the total
energy of a stellar population comes from PMS stars (Figure~6).
\begin{figure}
 \psfig{figure=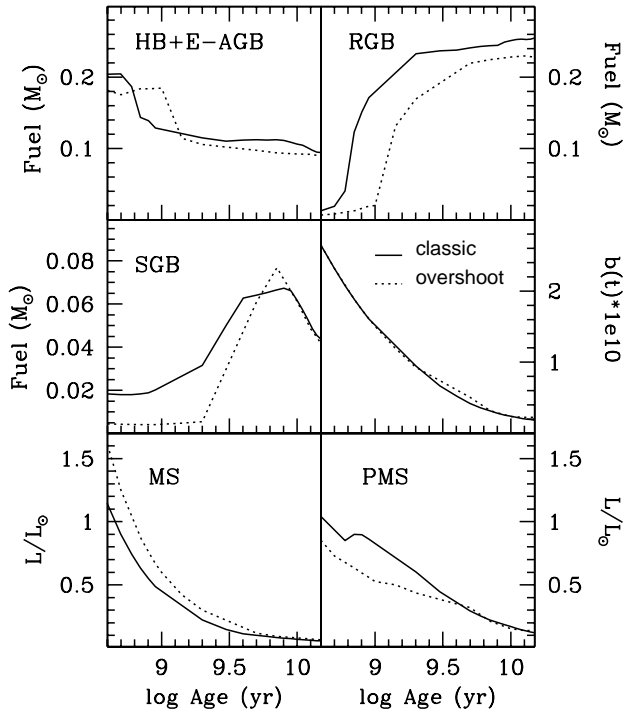,width=\linewidth}
  \caption{Influence of the stellar tracks. Comparison between the EPS
  ingredients obtained with classical (solid line, from Cassisi et
  al.) and overshooting tracks (dotted line, from Girardi et al. 2000
  and Salasnich et al. 2000). From top left to bottom right the
  various panels show: the total fuel consumptions (in {\rm
  $M_{\odot}$}) for the various PMS phases, as a function of the age
  of the SSP; the evolutionary flux; the bolometric luminosity of MS
  and post-MS. For both tracks the metallicity is half-solar
  ($Z=0.008$), and only ages larger than 0.3 Gyr are plotted. The
  TP-AGB fuel is not considered for this comparison since it does not
  depend on the adopted tracks, as explained in the text.}
\label{fuelpadfra}
\end{figure}

The dependence of MS and PMS energetics on stellar evolutionary tracks
is illustrated in Figure~7. The transition from a not degenerate to a
degenerate helium core, that marks a well developed RGB phase and a
drop in the HB fuel, occurs at later epochs if overshooting is taken
into account in the stellar models.  This is shown in the top panels
of Figure~\ref{fuelpadfra}, where the fuels obtained with classical
(solid) and overshooting (dotted) tracks as functions of the age of
the stellar populations are compared. The RGB phase transition occurs
at $0.7~\rm Gyr$ in classical models and at $1~\rm Gyr$ in models with
overshooting.  

The age marking the onset of the RGB phase transition was early
recommended by Barbaro \& Pigatto (1984) as a suitable observational
check of the overshooting hypothesis. This test has been recently
performed by Ferraro et al.~(2004), where deep, VLT-based, $JHK$ CMDs
of LMC GCs allow the quantification of the population ratios between
RGB and He-clump, as functions of the GC age. The comparison with the
theoretical predictions of classical and overshooted tracks (the same
tracks shown in Figure~\ref{fuelpadfra}) clearly exclude the latter,
in that they predict a substantially later development of the RGB (see
Figures~8,9 in Ferraro et al.~2004). This result supports our use of
classical evolutionary tracks for our standard SSP models.
Figure~7 (bottom left-hand panel) also shows that the MS luminosity of
the tracks with overshooting is larger than that of the classical one,
while the trend is reversed for the PMS (bottom right-hand panel), for
the reasons already explained. Note that the evolutionary fluxes are
very similar (central right-hand panel), which is mostly due to the
similarity of the time derivative of the turnoff masses. Finally, the
SGB phase appears also to be a little delayed (central left-hand
panel), but its impact on the results is very small.

Because of the earlier RGB phase transition, at $t\sim~1~\rm Gyr$ the
SSPs based on the Cassisi's tracks are nearly 40 per cent more
luminous than those based on the Padova tracks. An output that is
affected by these variations is obviously the $M/L$ ratio, a key
quantity used to determine the stellar mass of galaxies (see
Section~\ref{ml}).
\subsection{Temperatures}
\label{temperatures}
The second matrix contains the distribution of effective temperatures
and surface gravities of the evolutionary phases $j$ for the various
ages and metallicities. As already mentioned, every evolutionary phase
is split into a certain number of so-called {\it photometric
sub-phases}, inside which the spread in effective temperatures $\teff$
is $\lapprox~100~{\rm K}$. This was found to be appropriate for a good
tracing of the varying spectral type along a phase\footnote{It
should be also noted that model atmospheres are provided with $\Delta
\teff\sim200~{\rm K}$.}.
\begin{figure}
 \psfig{figure=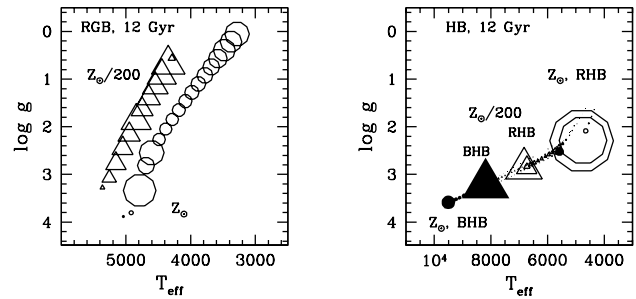,width=\linewidth}
  \caption{Relation between fuel consumption and temperature
  subphases. The left-hand panel shows the subphases along the RGB of
  12 Gyr SSPs with solar and $\zsun/200$ metallicities (circles and
  triangles, respectively). The right-hand panel shows the subphases
  along the HB for the same metallicities, filled and open symbols
  refer to blue and red horizontal branch morphologies,
  respectively. In both panels the symbol sizes scale with the fuel
  consumptions. The big circles in the left-hand panel correspond to
  the position of the so-called RGB bump.}
\label{tesub_fuel}
\end{figure}
What is revelant to evolutionary synthesis models is that the fuel
consumption is evaluated specifically in each subphase. In general the
consumption of energy is not homogeneous with temperature. This is
visualized in Figure~\ref{tesub_fuel}, where the temperature/gravity
subphases for old (12 Gyr) RGBs and HBs are displayed for two
metallicities (circles for solar; triangles for $\zh=-2.25$), with the
symbol size being proportional to the fuel. The fuel consumption along
the RGB of a metal-rich stellar population is enhanced at the
so-called RGB {\it bump} (big circles in the left-hand panel of Figure
~\ref{tesub_fuel}), when the H-burning shell reaches the internal
layer that was previously mixed through the first convective
dredge-up, gets fresh fuel and therefore spends a longer time in this
location (Sweigart, Greggio \& Renzini~1989). The RGB {\it bump} is
rather close to the He clump. If the total RGB fuel would be assigned
homogeneously along the RGB track, the weight of the {\it bump} would
be unappropriately distributed along the whole track, in particular
would be given to the tip, with the effect of overestimating the
near-IR flux of the SSP. The fuel consumption approach implemented
here where the evolutionary timescale is considered is an efficient
way of taking the bump into account. The bump is predicted to almost
disappear at decreasing metallcity (Sweigart, Greggio, Renzini~1989),
and indeed the RGB fuel in the metal-poor stellar population results
to be rather homogeneously distributed (triangles in
Figure~\ref{tesub_fuel}). Note that the significant contribution of
the RGB tip to the fuel comes from the high luminosity of that
subphase.

The right-hand panel of Figure~\ref{tesub_fuel} illustrates the effect
of the Horizontal Branch morphology on the temperature of the
subphases. For the same ages and metallicities two options for the HB
morphology are shown, red (open symbols) and blue (filled symbols). We
remind that blue/red HBs mean that the whole HB lifetime is spent on
the blue/red side of the RR-Lyrae strip, while intermediate HB is used
to refer to the mixed cases\footnote{Traditionally the separation
between red and blue HBs was assigned on the basis of the location of
the RR-Lyrae stars in metal-poor Milky Way GCs (Dickens~1972), whose
typical temperature is $\teff\sim 7000~\rm K$. Blue Horizontal
Branches (BHBs) are those in which most of the stars are found at
temperatures hotter than $\teff_{RR-Lyrae}$, Red Horizontal Branches
(RHBs) when most of the stars lie at cooler temperatures, Intermediate
Horizontal Branches (IHBs) for the mixed cases.} In order to trace
properly the HB evolution, we use the evolutionary track for the
helium burning phase of the mass that is obtained after mass-loss is
applied to the RGB track (see Section~\ref{HB}). In the HB phase, most
fuel consumption occurs on the so-called Zero Age Helium Burning
(ZAHB) that usually corresponds to a very narrow temperature range,
the evolution from the tip-RGB to the ZAHB and to the ZAHB to the
E-AGB happening on very short timescales ($\sim 1$ Myr). However in
presence of strong mass loss like in the case of a BHB at high $Z$
(filled circles) the evolutionary timescale can be significant at
various temperature locations.

Stellar effective temperatures depend crucially on the efficiency of
convective energy transfer, parametrized by the mixing length
parameter ${\alpha}$. The latter cannot be derived from first
principles, and as far as we know it could be connected to several
stellar parameters, such as the stellar mass, the evolutionary status
or the metallicity, and its calibration with observational data is
certainly required. The use of uncalibrated theoretical effective
temperatures in EPS computations is extremely dangerous, as
${\alpha}$ is the synchronization of the EPS clock (RB86).

The calibration of the mixing-length for the Cassisi's tracks is
described in Salaris \& Cassisi (1996). The tracks with solar
metallicity are computed for ${\alpha}=2.25$, a value that matches the
Sun. This same value is kept in the tracks with supersolar
metallicities (Bono et al.~1997). At sub-solar metallicities instead
the mixing-length parameter is not assumed to be the same, but to vary
with $Z$, such that the temperatures of the RGB tips of Milky Way GCs
are reproduced. The values range from 2 to 1.75, with a trend of
decreasing mixing-length with metallicity (see table~2 in Salaris \&
Cassisi~1996). In the Girardi et al.~(2000) tracks from the Padova
database instead the same value of the mixing-length parameter that is
calibrated with the Sun is assumed at all metallicities.
\begin{figure}
 \psfig{figure=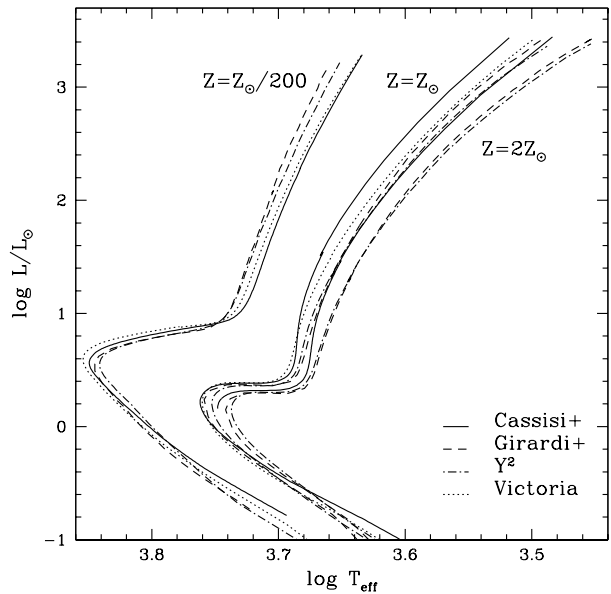,width=\linewidth}
  \caption{Influence of the stellar models on RGB temperatures. For
  metallicities $\zsun/200, \zsun, 2 \zsun$, classical (solid line,
  from Cassisi et al.) and overshooting tracks (dotted line, from
  Girardi et al. 2000) are shown up to the RGB tip. The Padova tracks
  of Girardi et al.~2000 have cooler RGBs at high metallicity (solar
  and above) and warmer ones at low metallicities. At least part of
  this effect should arise from the adopted mixing-length
  parameter. Also plotted are the isochrones from the Victoria group
  (dotted line, Vandenberg et al.~{\it in preparation} for solar
  metallicity; Bergbusch \& Vandenberg~(2000) for sub-solar
  metallicities) and from Yale ($\rm Y^{2}$, dashed line, Yi et
  al.~2003). All models have the age of 10 Gyr.}
\label{tergbtracks}
\end{figure}

The resulting differences in the RGB temperatures are shown in
Figure~\ref{tergbtracks}, where the RGBs of old (10 Gyr) isochrones
from Cassisi et al. (solid lines) and from Padova (Girardi et
al.~2000, dotted lines) are shown for three metallicties ($\zsun/200,
\zsun, 2~\zsun$, from left to right). At high metallicities, the
Padova RGBs are cooler than those of the Cassisi's tracks, in such a
way that the 2~$\zsun$ RGB of Cassisi coincides with the solar one by
Girardi et al. Since the cooler temperatures are proper to the whole
RGB, and not only to the tip, for what said before (see also
Figure~\ref{tesub_fuel}) it has to be expected that the optical/IR
flux ratio of metal-rich SSP models will depend on the choice of the
tracks. This effect will be quantified in
Section~\ref{uncertainties}. Unfortunately, as already mentioned, at
these high metallicities, that are the most relevant to massive
galaxies, the calibration of the models is hampered by the lack of
metal-rich GCs with independently known ages and metallicities. The
only two objects useful to this purpouse are the two metal rich GCs of
the Baade window (NGC~6553 and NGC~6528, Ortolani et al.~1995), whose
total metallicity is around solar (Thomas, Maraston \&
Bender~2003a). The complication here is that these two objects have
enhanced $\afe$ ratios (Barbuy et al.~1999; Cohen et al.~1999),
therefore the proper calibration requires the use of stellar tracks
accounting for this effect. Element ratio effects are being
incorporated in stellar models (e.g. Bergbusch \& Vandenberg~1992,
2001; Salasnich et al.~2000; Kim et al.~2002), but for the stellar
tracks considered here, their $\afe$-enhanced version was either not
yet available at the time these models were computed (Cassisi's
tracks) or does not appear to be convincing (see Thomas \&
Maraston~2003 for the Salasnich et al.~2000 tracks).

At lower metallicities the opposite effect is found, i.e. the Cassisi
tracks appear to have cooler RGBs' than the Padova ones. At least part
of the effect must originate from the treatment of the
mixing-length. However the metodology by Salaris \& Cassisi~(1996)
depends necessarily on temperature/colour transformations, as the RGB
theoretical temperature luminosity relation must be converted into the
observed colour-magnitude diagram. Therefore it is hard to push any
strong conclusion on which parametrization of the mixing-length is
better.  Generally, any calibration of stellar tracks depends upon the
adopted temperature/colour transformations, and in particular for the
coolest part of the RGB these are notoriously uncertain.

The comparison with the Padova tracks is the most relevant to
evolutionary population synthesis issues, since these tracks are
adopted by all existing EPS codes except the one presented
here. However several other isochrones exist in the literature. In
Figure~\ref{tergbtracks} the latest Yale models ($Y^{2}$, Yi et
al.~2003) and the models of Victoria (D.A. Vandenberg, {\it private
communication} for solar metallicity; Bergbusch \& Vandenberg ~(2000)
for the subsolar metallicity) are shown. Both sets of tracks include
overshooting. The $\rm Y^{2}$ isochrones (dot-dashed) behave very
similarly to the Padova ones at all metallicities. At solar metalicity
the Victoria models (dotted line) agree quite well with the Cassisi's
one till the early portion of the RGB, after which the track departs
toward cooler temperatures reaching values around the tip that are
more similar to those of the Padova or Yale isochrones. Interestingly
the metal-poor isochrone of Victoria has a RGB rather similar to that
of Cassisi, which suggests that the calibration of the mixing-length
is not the whole story. At supersolar metallicities the Victoria
isochrones are not available. In Section~\ref{uncertainties} we will
show the impact of the various RGBs' on the integrated colours of
SSPs.
\subsection{Transformations to observables}
\label{transf}
The third matrix contains the transformations to the observables, used
to convert effective temperatures and surface gravities into a
spectral energy distribution (SED). The transformations can be either
theoretical or empirical. The models of M98 used a mix bag of
ingredients. They rely on the classical Kurucz (1979 and revisions)
model atmospheres for $3500\lapprox \teff (\rm K) \lapprox 35000$,
complemented with models for M giants by Bessel et al.~(1989) for
cooler temperatures, and with empirical colours for TP-AGB stars (see
M98 for references). The models presented here are partially revised
in this respect because the ingredients of the mixed matrix
constructed by M98 have been in the meantime compiled into one library
(see Section~\ref{atmo}), in which border effects between the merged
libraries are considered in far more detail than it was done in
M98. Also the empirical ingredients for TP-AGB stars have been
revised, because of the availablity of complete SEDs
(Section~\ref{tpagbsed}).
\subsubsection{Model atmospheres}
\label{atmo}
The synthetic stellar spectra are taken from the spectral library
compiled by Lejeune, Cuisinier and Buser (1998, in its latest version
as available on the web, hereafter the BaSel library). This library
has become widely used in population synthesis studies and was
obtained by merging the Kurucz library of model atmospheres ({$\teff
\gapprox 3500~{\rm ^oK}$) with model atmospheres for cooler stars (for
references and full details see Lejeune et al.). As in M98, a
quadratic interpolation in ($\teff,logg$) is performed on this library
to compute the stellar spectra appropriate to each subphase. The BaSeL
library is provided for various iron abundances. We obtain the
appropriate one to each set of tracks by interpolating linearly in
$\feh$. Finally, a blackbody spectrum is assigned $\teff \gapprox
50,000~{\rm K}$.
The use of the BaSel library allows the computation of model SEDs with
low spectral resolution, i.e. 5-10~\AA up to the visual region, 20 to
100~\AA~in the near-IR. Bruzual \& Charlot~(2003) adopt STELIB (Le
Borgne et al.~2003, see also Le Borgne et al. 2004), an empirical
spectral library of stars with metallicity around solar and a much
higher spectral resolution (3~\AA), to compute a set of SSPs with
solar metallicity. They compare the integrated colours $B-V$ and $V-K$
obtained with both BaSel and STELIB. The colour differences as induced
by the spectral transformations result to be at most a few hundreths
of magnitude in a wide range of ages.
\subsubsection{Empirical spectra for TP-AGB stars}
\label{tpagbsed}
As is well known, current synthetic spectral libraries do not include
spectra for Carbon-rich and Oxygen-rich stars populating the TP-AGB
phase, although some theoretical computations begin to be available
(Lloidl et al.~2001). For the spectra of this type of stars we use the
empirical library by Lan\c {c}on \& Mouchine~(2002). The latter is based on
the library of individual stellar spectra by Lan\c {c}on \& Wood~(2000),
that collects observations of C-,O-type stars in the Milky Way and
Magellanic Clouds. Since individual stellar spectra of such cool,
variable stars are subjected to strong star-to-star or
observation-to-observation variations, Lan\c {c}on \& Mouchine~(2002) have
constructed mean templates of the Lan\c {c}on \& Wood~(2000) library,
obtained by averaging observations of individual stars. In this work
we will use the average templates.
\subsection{Recipes for critical stellar phases: TP-AGB and HB}
\label{recipes}
\subsubsection{Mass loss in red giants}
\label{massloss}
The stellar temperatures and luminosities during those evolutionary
phases that follow episodes of, or suffer themselves of, stellar mass
loss, cannot be predicted by stellar tracks. This comes from the fact
that a theory relating mass loss rates to the basic stellar parameters
does not exist. Therefore mass loss has to be parametrized and its
efficiency be calibrated with data. Due to such complication we call
these stellar phases `critical'. The amount of mass loss is usually
parametrized by means of the Reimers (1977) empirical formula
$dM=-\eta \cdot (L/(logg \cdot R))\cdot dt$, where $M$, $L$, logg, $R$
stand for mass, emergent luminosity, surface gravity and radius,
respectively, of the stellar configuration in its lifetime $dt$. The
parameter $\eta$ introduced by Fusi Pecci \& Renzini (1976) takes into
account the efficiency of mass-loss. 
%Mass loss affects the evolution
%of massive stars ($M \gapprox 10~ \rm \msun$) already on the Main
%Sequence. 
In stars of intermediate mass ($2\lapprox M/\msun\lapprox
8$), mass loss influences significantly their post Main Sequence
evolution, i.e. the energetics and the temperature, along the TP-AGB
(Section~\ref{TP-AGB}). In low-mass stars ($M \lapprox 2~\rm \msun$)
mass-loss occurs also during the RGB, particularly towards the tip,
affecting the subsequent HB evolution and hence the HB morphology
(Section~\ref{HB}).

It is important to notice that the efficiency $\eta$ cannot depend too
much on metallicity (Renzini~1981), because the observed HB morphology
of GCs is almost always red at high-metallicity, implying that
mass-loss does not increase significantly with the metal
abundance. Renzini~(1981) evaluates $\eta\propto Z^{\rm x}, {\rm
x}\lapprox 0.2$.

In the following sections we describe the modelling of HB and TP-AGB.
\subsubsection{The Horizontal Branch morphology}
\label{HB}
The amount of mass loss during the RGB phase is computed by
integrating the Reimers formula along the RGB track. The efficiency
$\eta$ was calibrated by Fusi Pecci \& Renzini (1976) by comparing the
mass of RR-Lyrae in the instability strip of the metal-poor MW GC M13
($\zh\sim-1.5$), with its turnoff mass, and found to be
$\sim0.33$. This value of $\eta$ is by definition appropriate only at
this metallicity and for the RGB tracks used by Fusi Pecci \&
Renzini~(1976) and it has to be re-obtained for other chemical
compositions and when other stellar models are used.

The approach followed by Maraston \& Thomas~(2000) was to compute the
integrated $\hb$ line, that is very sensitive to the HB morphology
(see also de Freitas Pacheco \& Barbuy~1995; Poggianti \&
Barbaro~1997; Lee et al.~2000) and to compare it with galactic GCs of
known ages, metallicities and HB morphologies. The value of $\eta$
appropriate to $\zh\sim-2.25$ was found to be 0.2.

The procedure of using one value of $\eta$ per metallicity (and age)
aims at recovering the average trend of a bluer HB morphologies with
decreasing metallicity, the latter being the 1-st parameter ruling the
HB morphology. The trend is easy to understand in terms of stellar
evolution. At low metallicity the evolutionary mass is smaller at
given MS lifetime (because the stars are more compact and hotter and
the nuclear burning more efficient), therefore the production of
hotter effective temperatures by envelope removal is easier (helped
also by the lower metal content). However, as well known a large
scatter is found in the HB morphology of GCs with the same nominal
metallicity, a still unexplained fact that is recalled as the 2-nd
parameter effect. The account of all possible HB morphologies in a SSP
model is not useful, but it is sensible to provide models with a few
choices for the HB morphology, that are able to encompass the observed
average trend and scatter.

\begin{figure}
 \psfig{figure=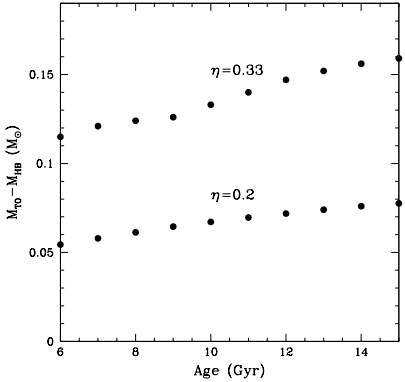,width=0.7\linewidth}
  \caption{The amount of mass loss along the RGB, expressed as the
  mass difference between the turnoff mass and the mass at the Helium
  burning stage, as function of age for two mass loss efficiency
  $\eta$.}
\label{mloss}
\end{figure}
The adopted value of $\eta$ are $\eta=(0.2;0.33)$ at $\zh=(-2.25;
-1.35)$. The correspondent amount of mass loss along the RGB is shown
in Figure~\ref{mloss}, where the mass difference (in $\msun$) between
the turnoff mass and the mass in the HB phase is shown as function of
age for the two adopted $\eta$'s. These values are in remarkable
agreement with those computed by Greggio \& Renzini~(1990) using a
different set of evolutionary tracks.
\begin{figure}
 \psfig{figure=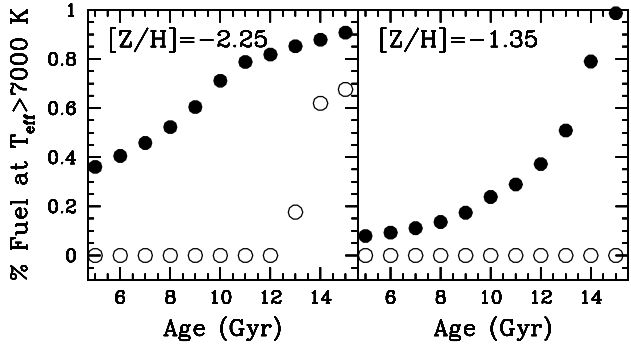,width=\linewidth}
  \caption{The HB morphology as function of age and metallicity that
  results from having applied $\eta=0.2$ and 0.33 at $\zh=-2.25$ and
  $\zh=-1.35$, respectively (filled symbols). The $y$-scale gives the
  percentage of the total HB fuel that is spent at
  $\teff\gapprox~7000~\rm K$.  The open symbols show the resulting
  fuel partition when no mass loss is applied during the RGB evolution.}
\label{hbmor}
\end{figure}
The resulting HB morphology is shown in Figure~\ref{hbmor} by means of
the percentage of fuel burned at $\teff \gapprox 7000$ K, according to
the definition of BHB given in Footnote 1, Section~\ref{temperatures}.
The HB morphology is almost completely blue (BHB) at $\zh=-2.25$ and
$t\gapprox 10$ Gyr, and at $\zh=-1.35$ and $t\gapprox 14$ Gyr. The
morphology is red (RHB) at ages lower than 6 Gyr for $\zh=-1.35$,
intermediate (IHB) otherwise. At $\zh=-2.25$, the HB morphology is
never completely red, because the metallicity is so low that the HB
track spends a no negligible amount of fuel bluewards the RR-Lyrae
strip even without mass-loss (open symbols in Figure~\ref{hbmor}). The
effect of age is such that the higher the age is, the lower is the
stellar mass and the higher is its effective temperature for the same
mass loss. Therefore at a given $\eta$ the HB morphology gets bluer
with increasing age (see also Lee et al.~2000). The SSP models
computed with these choices of $\eta$ have been proven to give a very
good match of the $\hb$ line and the mid-UV spectra (Maraston \&
Thomas~2000) and of the higher-order Balmer lines (Maraston et
al.~2003) of Milky Way GCs as function of their total metallicities.

At higher metallicities, the HB of Milky Way GCs is found to be red in
almost all cases. However, we know at least two examples of metal-rich
GCs that have extended HB morphologies. These are the Bulge GCs NGC
6441 and GC 6388 (Rich et al.~1997) with $\zh\sim-0.55$. As shown in
Maraston et al.~(2003), the models by Maraston \& Thomas (2000) were
able to reproduce the observed Balmer lines of these two clusters by
assuming moderate mass-loss along the RGB, implying 10-15 per cent of
fuel to be spent blueward the RR-Lyrae strip. This fraction is
consistent with the observed value.

The finding of Rich et al.~(1997) calls for the need of models with
BHBs also at high-metallicities. For this purpouse we seek the $\eta$
value at which the HB fuel is spent (almost) entirely blueward the
RR-Lyrae strip. For the Cassisi tracks we find
$\eta=(0.85,0.45);(1.0,0.7);(0.94,0.66)$ for 10 and 15 Gyr and
metallicities half-solar, solar and twice-solar, respectively. The
masses at the HB phase are then $\sim 0.5\div0.55$, whose temperatures
reach $\sim~9000~{\rm K}$. This amount of mass-loss means to remove
nearly half of the initial stellar mass during the RGB.  Note finally
that by choosing a larger $\eta$, one gets an even bluer HB
morphology. The SSPs with high-Z and BHB are computed for the ages of
10 and 15 Gyr.

As a last remark, the variation of the HB fuel between models with and
without mass-loss is of having less fuel in the models in which mass
loss is applied, that stems from the lower evolutionary mass (see
Figure~\ref{fuelfasi}). Such differences amount to roughly 7 per cent
at $\zh=-2.25$ , to 15 per cent at $\zh=-1.35$, to 27 per cent at
half-solar metallicity, to 40 per cent at solar metallicity and to 54
per cent at $2~\zsun$.
\subsubsection{The Thermally-Pulsing AGB: inclusion in the 
integrated spectra}
\label{TP-AGB}
M98 present SSPs in which the TP-AGB phase was included
semi-empirically in the models, using a table of theoretical fuel
consumptions (from Renzini~1992) and calibrating them with
measurements of the bolometric contribution of the TP-AGB phase to the
total light in intermediate age LMC GCs (from Frogel, Mould \&
Blanco~1990, see Figure~3 in M98). Basically, the observed
contributions fix the left-hand side member of Eq.~\ref{lpms} written
for $j=$TP AGB, which allows the evaluation of the fuel, using the
evolutionary flux appropriate for the given age. This calibration of
the fuel equals to determine empirically the mass-loss efficiency
along the TP-AGB phase, which is found to be $\eta\sim1/3\div2/3$.

Maraston et al.~(2001) extend the SSP models to half and twice solar
metallicity, introducing an analytical recipe that connects the amount
of TP-AGB fuel to the envelope mass at the first thermal pulse
(beginning of the TP-AGB phase). Briefly recalling, the fraction of
envelope mass that is burned as TP-AGB fuel is determined for the
turnoff masses at solar metallicity from the fuel table of M98. As a
next step, the relations between envelope and total masses for the
turnoff stars with different chemical compositions (taken from the
stellar tracks at the termination of the E-AGB) are used to establish
which fractions of them goes into TP-AGB fuel as function of
metallicity. In this scaling it is assumed that the mass-loss
efficiency (which is derived from the Magellanic Clouds GCs) does not
depend on metallicity. As a result of this assumption together with
the fact that the turnoff mass is rather insentive to metallicity, the
total TP-AGB fuel does not depend too much on the chemical composition
(Maraston et al.~2001, see Figure~\ref{fuelfasi}). This is a robust
result for two reasons. 1) Mass-loss is not affected significantly by
metallicity, as we discussed in the previous section. 2) The TP-AGB
fuel decreases very little with increasing mass-loss efficiency.
Renzini \& Voli~(1981) show that by doubling $\eta$ from 0.33 to 0.66
the fuel of a $2 \msun$ star with solar metallicity decreases by only
20 per cent.

Instead the metallicity influences the partition of the total fuel
between C and M stars, according to the rationale of Renzini \&
Voli~(1981, adopted in both M98 and Maraston et al.~2001). Briefly
summarizing, in metal-poor chemical compositions the abundance of
oxygen in the envelope is lower and a lower amount of carbon has to be
dredged-up in order to reach ${\rm C/O}>1$, hence binding the whole
oxygen into CO molecules (those with the highest binding energy). The
residual carbon is then available to produce CH, CN, C$_{2}$ molecules
and carbon stars are made. Therefore a metal-poor stellar population
is expected to have more carbon stars than a metal-rich
one. Quantitatively, the fuel in C-stars doubles by halving the
metallicity.
\begin{figure}
 \psfig{figure=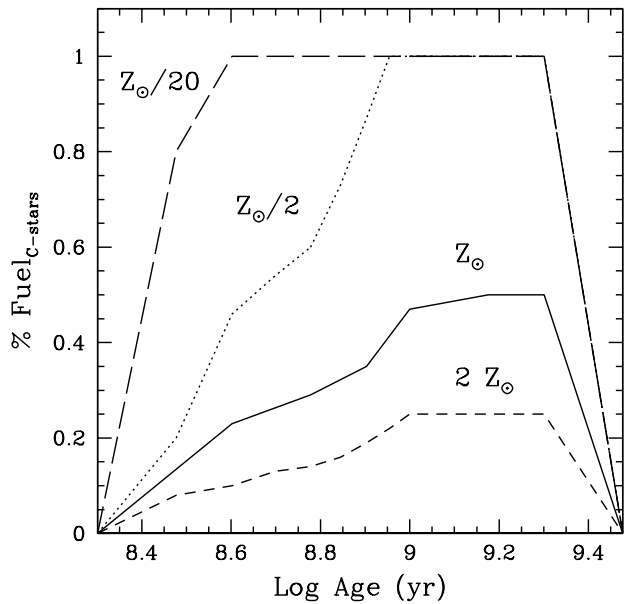,width=\linewidth}
  \caption{The percentage of the TP-AGB fuel spent in Carbon star mode
  for the various metallicities, as a function of age.}
\label{CvsM}
\end{figure}
The adopted recipe is given in Figure~\ref{CvsM}, where the percentage
of the TP-AGB fuel that is spent by Carbon-type stars as a function of
metallicity is shown. The reference metallicity to which the scaling
of Renzini \& Voli~(1981) is applied, is the $0.5~\zsun$ metallicity
(dotted line), because this is the chemical composition of the
Magellanic Clouds GCs that are used as calibrators\footnote{Frogel,
Mould \& Blanco~(1990) measure also the luminosity contributions of C
and M-type stars as functions of the GC age.}  Solar and twice solar
metallicities (solid and short-dashed line, respectively) are assigned
halves and one quarter of the fuel in C-star of the $0.5~\zsun$
metallicity, respectively\footnote{ L. Greggio and M. Mouchine pointed
out that the percentage of fuel spent as C and M stars in Table~4 and
Figure~2 of Maraston~(1998) is reversed. This mistake concerns only
the graphics and not the models. In any case the reader is referred to
the present Figure~12 and to the tables provided at the model WEB
page.}. At $\zh\sim-1.35$ (long-dashed line) the fuel in C-star is a
factor 10 larger than at $0.5~\zsun$, therefore the TP-AGB fuel is
almost always spent by C-stars. Finally, at $\zh\sim-2.25$ the whole
TP-AGB fuel is assigned to C-stars. These numbers can be compared to
data of resolved C-stars in galaxies and help constraining the age
distribution of the stellar populations. For example, Davidge ~(2003)
finds that the C-star component is 10 \% of the whole AGB in the dwarf
galaxy NGC~205. According to Figure~12 this implies that either a
burst has occurred less than $\sim$ 200 Myr ago (as favoured by
Davidge 2003) or the stellar population is globally old. Also
interesting is the finding of a conspicuous population of C stars in
the arms of M33 (Block et al.~2004), which could help constraining the
star formation history of this galaxy.

Note that the C/M ratio as a function of metallicity by Renzini \&
Voli~(1981) is confirmed by the recent TP-AGB models of Marigo et
al.~(1999), as inferred from their figures since quantities are not
tabulated.

Our previous models were restricted to broad-band colours due to the
unavailability at the time of spectra, either theoretical or
empirical, appropriate to TP-AGB stars. However, in order to be useful
for high redshift studies, the TP-AGB phase has to be included in the
synthetic spectral energy distribution. This improvement of the models
is performed here.  The fuel is distributed among the empirical
spectra of the individual C-,O-type stars by Lan\c {c}on \&
Mouchine~(2002). The spectral type that starts the TP-AGB phase is
chosen to be close to that of the subphase terminating the E-AGB
phase. Spectral types of both O-and C-stars are then included, as
described above. 
\section{Results}
\label{results}
\begin{figure*}
 \psfig{figure=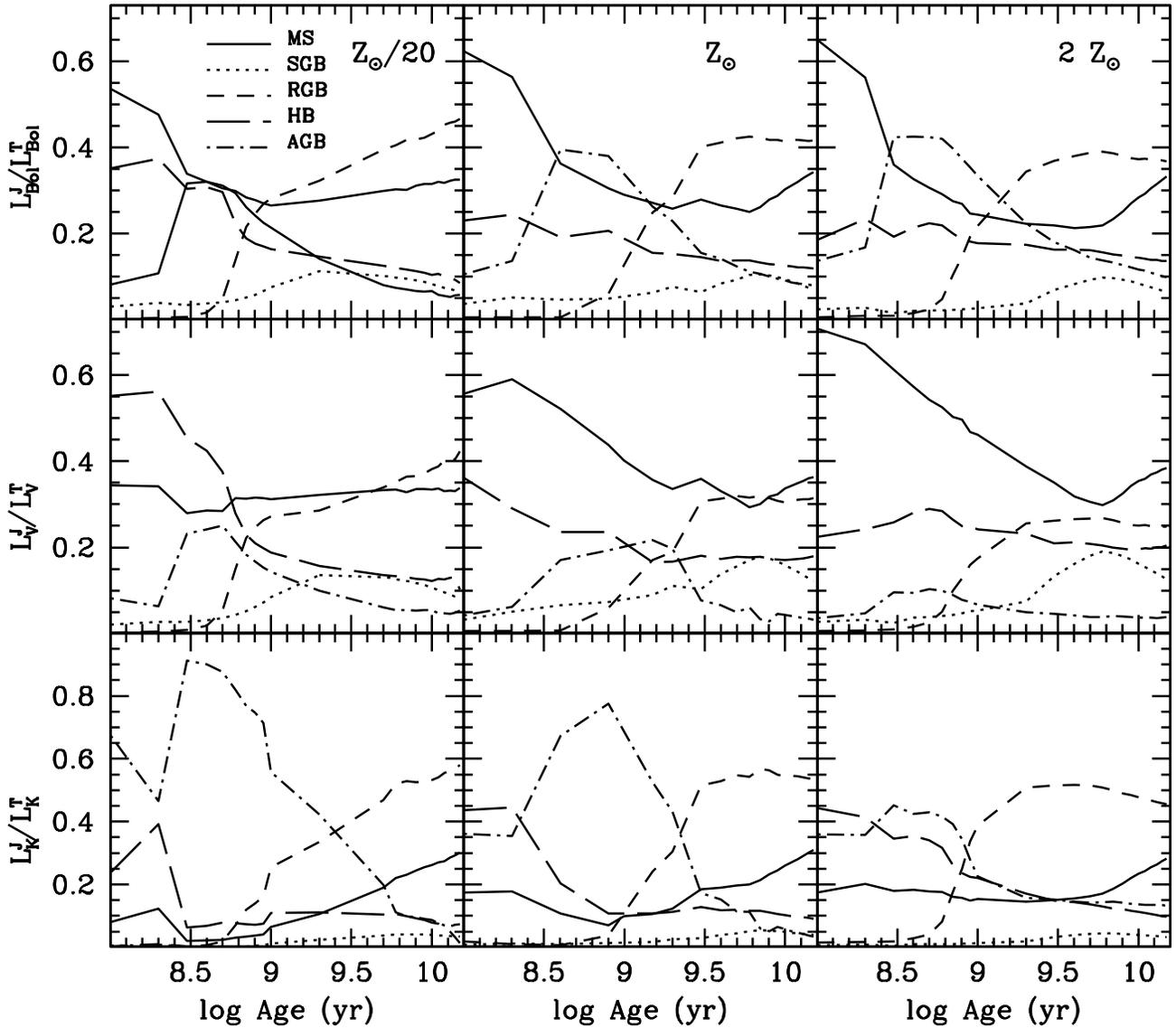,width=\linewidth}
  \caption{Luminosity contributions in bolometric, $V$ and $K$ (from
  top to bottom) of evolutionary phases and their dependences on age
  and metallicity. From left to right metal-poor, solar metallicity
  and metal-rich SSPs (with a Salpeter IMF) are shown,
  respectively. Note that the $y$-scale of the bottom panel is not the
  same as in the other two panels.}
\label{contrpha}
\end{figure*}
The model output are the integrated spectrophotometric properties of
SSPs, namely spectral energy distributions (SEDs), broad-band colours
(in Johnson-Cousins, SLOAN, HST filter systems), spectral line
indices, mass-to-light ratios, bolometric corrections, etc. These are
functions of the main parameters of the SSP model: the age $t$ and the
chemical composition $X,Y,Z$ (refereed to as $\zh$\footnote{The
notation \zh~is used to indicate total metallicities, i.e. the total
abundance of heavy elements with respect to hydrogen normalized to the
solar values, i.e. $\zh=\log({{\rm Z}/{\zsun}})-\log({{\rm H}/
\hsun})$. By $\feh$ we mean the abundance of iron with respect to
hydrogen normalized to the solar values, i.e. $\feh=\log({{\rm
Fe}/{\fesun}})-\log({{\rm H}/ \hsun})$. If elements have solar
proportions then $\feh=\zh$. In case of $\alpha$-element enhancement,
the relation between $\feh$ and $\zh$ is: $\feh=\zh-0.93*(\afe)$
(Thomas, Maraston \& Bender~2003).}). Ages and metallicities of the
SSP grid are listed in Table~\ref{SSPgrid}. Models are generally given
in time steps of 1 Gyr, for ages larger than 1 Gyr, and of 0.1 Gyr for
smaller ages. The Table indicates also the source for the input
stellar tracks/isochrones, according to the description and references
given in Section~\ref{energetics}. Model assumptions concern the
initial mass function (IMF) and the morphology of the Horizontal
Branch. The models are computed for two choices of the IMF, namely
Salpeter~(1955) and Kroupa~(2001), and are provided for two different
assumptions regarding the mass loss along the RGB
({Section~\ref{HB}}), which result into various HB morphologies.
\begin{table}
 \centering
  \caption{\label{SSPgrid}
  Ages, metallicities and input tracks of the SSP grid}
 \begin{tabular}{@{}rccc}
  & & & \\
 $t (\rm Gyr)$ & $(Y,Z)$        & $\rm [Z/H]$  & $\rm Stellar~tracks$ \\  
  & & &  \\
     1--15   &  $0.230,10^{-4}$ & $-2.25$  & Cassisi \\
   $3\cdot 10^{-6}$--15   &  $0.230,0.001$  & $-1.35$  & Cassisi + Geneva\\
   $3\cdot 10^{-6}$--15  &  $0.255,0.01$   & $-0.33$  & Cassisi + Geneva\\
   $3\cdot 10^{-6}$--15  &  $0.289,0.02$  &  $+0.00$ & Cassisi + Geneva\\
   $3\cdot 10^{-6}$--15   &  $0.340,0.04$ &  $+0.35$ & Cassisi + Geneva\\
   1--15     &  $0.390,0.07$ & $+0.67$ & Padova \\ 
   \end{tabular}
\end{table}
\begin{table}
 \centering
  \caption{\label{SSPHbmor}
HB morphologies of the SSP grid}
 \begin{tabular}{@{}ccccc}
 & &  \\
   ${\rm [Z/H]}$ & ${\rm Age~(Gyr)}$ & $\eta_{\rm mass loss}$ & ${\rm HB~morphology}$ & $\rm Ref.$ \\  
    & & & & \\
   -2.25 &  $<=12$   &  0.00 & RHB & RI-HB \\
   -2.25 &  $>12$   &  0.00 & IHB & RI-HB \\
   -2.25 &  $<10$   &  0.20 & IHB & IB-HB \\
   -2.25 &  $>=10$   &  0.20 & BHB & IB-HB \\
   -1.35 &  {\rm all}  & 0.00 & RHB & RI-HB \\
   -1.35 &  $10--13$  & 0.33 & IHB & IB-HB  \\
   -1.35 &  $14--15$  & 0.33 & BHB & IB-HB \\
   -0.33 &  {\rm all} & 0.00 & RHB & RI-HB or IB-HB \\
   -0.33 & 10,15 & 0.55 & BHB  & high-Z BHB \\
    0.00 & {\rm all} & 0.00 & RHB & RI-HB or IB-HB \\
    0.00 & 10,15  & 0.55 & BHB & high-Z BHB  \\
    0.35 & {\rm all} & 0.00 & RHB & RI-HB or IB-HB \\
    0.35 & 10, 15 & 0.55 & BHB & high-Z BHB \\
   \end{tabular}
\end{table}
These are indicated in Table~\ref{SSPHbmor}. In the following
subsections we discuss the various model output and their comparisons
with observational data and with similar models from the literature.
\subsection{Luminosity contributions by stellar phases}
\label{resbolometric}
Figure~\ref{contrpha} shows the contributions from the various
evolutionary phases to the total bolometric, $V$ and $K$ luminosities
(from top to bottom) of metal-poor, solar metallicity, and metal-rich
SSPs, respectively (from left to right).

As in the solar metallicity SSP already discussed in M98, most
bolometric energy is shared by the three main phases MS, TP-AGB and
RGB independent of metallicity (cf. Figure~\ref{fuelfasi}). The
energetics of young SSPs ($t\la$ 0.3 {\rm Gyr}) is dominated by MS
stars, that of SSPs with $t\ga$ 2 {\rm Gyr} by RGB stars. SSPs with
ages in the range 0.3 $\la t\la$ 2 {\rm Gyr} are dominated by TP-AGB
\footnote{Note that in the figure the total AGB contribution,
i.e. that from E-AGB plus TP-AGB, is plotted, but we know from
Section~\ref{energetics} that the AGB phase is dominated by the
TP-AGB} stars except in the metal-poor stellar population, where AGB,
MS and HB have similar contributions. For ages older than $\sim$ 6
Gyr, the MS contributions tend to rise. This effect is caused by the
MS integrated luminosity decreasing slower than $ b(t) $. As the total
fuel keeps nearly constant, the net result is a lower total luminosity
as mainly due to a lower PMS luminosity.

The main contribution to the $V$-luminosity comes always from MS
stars, except for the metal-poor SSP, where at young ages the HB phase
has a relatively larger contribution, and at very old ages the RGB is
the dominant phase for the $V$-luminosity, which is due to the warm
RGB temperatures at such low metallicity.

Finally, the light in the $K$-band (and in the near-infrared in
general) in the age range 0.3 $\la t\la$ 2 {\rm Gyr} is dominated by
TP-AGB stars at every metallicity, a r\^ole taken over by the RGB at
old ages. These results demonstrate the importance of including the
TP-AGB phase for a correct interpretation of rest-frame near-infrared
colours and luminosities of $1~\rm Gyr$ stellar populations
\subsection{Spectral energy distributions with TP-AGB}
The synthetic SEDs of old stellar populations, in particular the
Horizontal Branch morphology, have been compared with IUE data of GCs
in Maraston \& Thomas~(2000) up to the metallicity of 47 Tucanae. In
this section we focus on the most relevant features of the new model
SEDs, namely the inclusion of the spectra of C- and O-type TP-AGB
stars. We will compare the SEDs with a sample of Magellanic Clouds GCs
for which data in the whole spectral range from $U$ to $K$ are
available, and with models from the literature. Further comparisons
with both observational data and models using broad-band colours and
spectral indices will find place in following sections.

\subsubsection{Fingerprints of $\sim~1~{\rm Gyr}$ populations}
\label{fingerTP-AGB}
\begin{figure}
 \psfig{figure=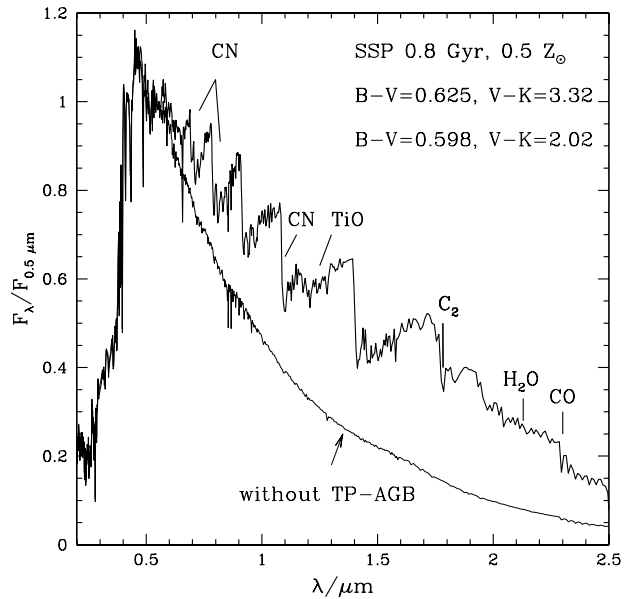,width=\linewidth}
  \caption{The effect of the TP-AGB phase on the spectral energy
  distribution of a 0.8 Gyr stellar population with $Z=0.5~\zsun$
  (thick line). The same SSP without the TP-AGB phase is shown as a
  thin line. The integrated $B-V$ and $V-K$ colours obtained on both
  SEDs are given. The most important absorption bands typical of
  C-stars (e.g. ${\rm C_2}$, CN), O-stars (e.g. $\rm H_{2}O$, TiO), or
  both (e.g. CO) are indicated. The relative proportions of fuel in C
  and O stars at this metallicity and age is 9:1
  (cf. Figure~\ref{CvsM}).}
\label{sedtp}
\end{figure}
The inclusion of the TP-AGB phase in a model SED is substantial at
ages in the range $0.3 \lapprox t/{\rm Gyr}\lapprox 2$ where the fuel
consumption in this phase is maximum (cf. Section~\ref{fuels}). This
is shown in Figure~\ref{sedtp}, where two model SEDs of 0.8 Gyr,
$0.5~\zsun$ with and without the TP-AGB phase are compared. While the
optical part of the spectra is insensitive to the presence of the cool
TP-AGB stars, the near-IR one changes dramatically. Not only the
absolute flux in the near-IR region increases by nearly a factor 3,
also peculiar absorption features appear (e.g. CN, $\rm C_{2}$ in
C-stars and ${\rm H_{2}O}$ and CO in O-stars, see e.g. Lan\c {c}on \&
Wood~2000 for more details). These absorption features besides the
integrated colors can be used as indicators of $\sim 1~{\rm Gyr}$
stellar populations in the integrated spectra of stellar systems. For
example, Mouhcine et al.~(2002) detected Carbon molecules absorptions
in the near-IR spectrum of W3, a massive GC of the merger-remnant
galaxy NGC 7252 that we previously suggested to be in the AGB phase
transition based on its $B,V,K$ colours (Maraston et al.~2001).
\begin{figure*}
 \psfig{figure=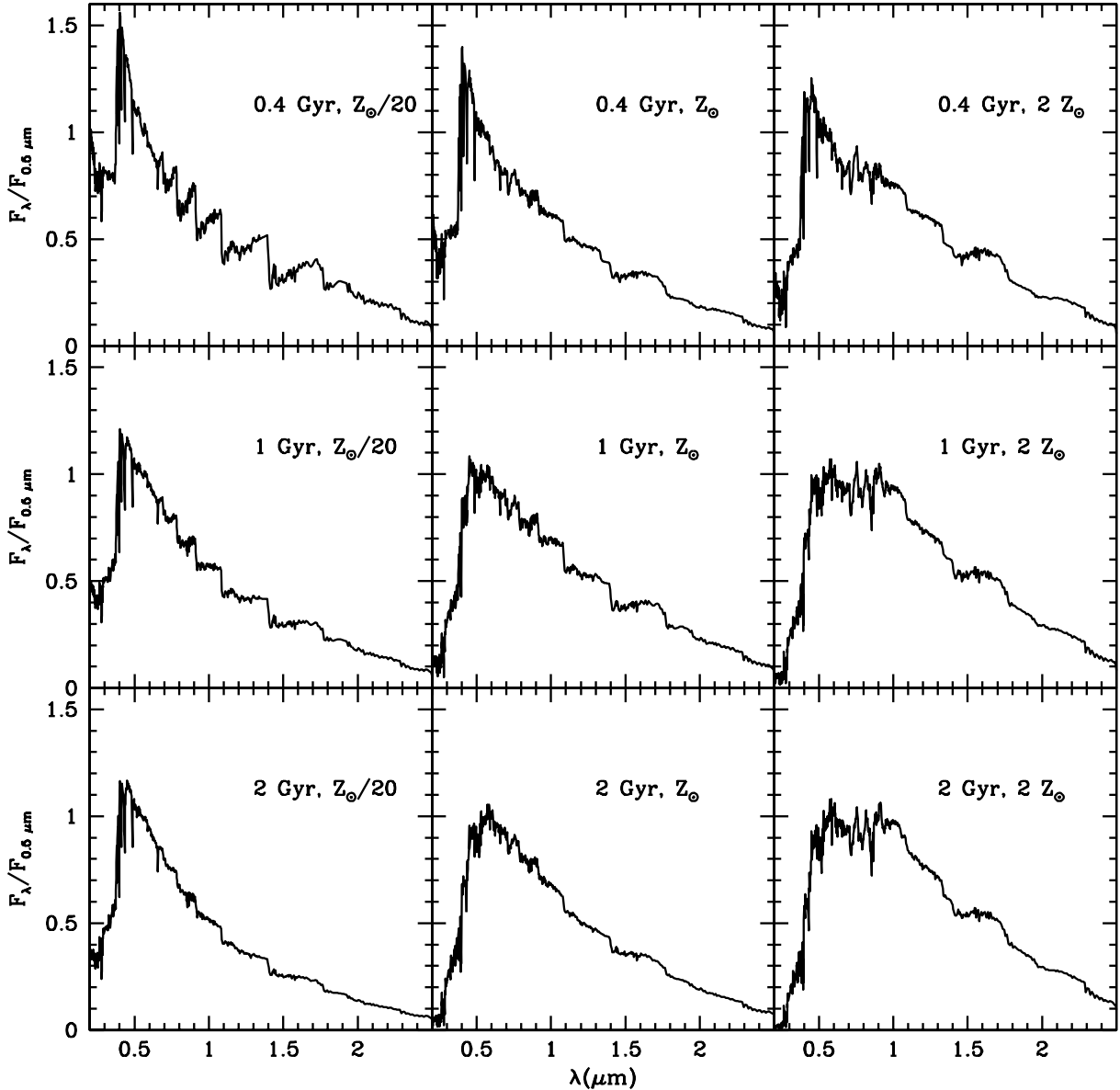,width=\linewidth}
  \caption{Effect of the metallicity of the TP-AGB stars on the SEDs
  of 0.4, 1 and 2 Gyr stellar populations. From left to right, the
  metallicity increases from $\zsun/20$ to $2~\zsun$, and with it the
  relative importance of Oxygen-rich stars over Carbon-rich stars.}
\label{sedtpZ}
\end{figure*}

As discussed in section~\ref{TP-AGB}, the relative proportions of C-
and O-stars depend on metallicity, for the model of Figure~\ref{sedtp}
the ratio being 9:1. The effect of considering other metallicities is
displayed in Figure~\ref{sedtpZ}, where 0.4, 1 and 2 Gyr SSPs with
chemical compositions $\zsun/20$, $\zsun$ and $2~\zsun$, are
shown. The larger total metallicity favours the production of O-rich
stars over that of C-rich stars. This is evident in the integrated
SEDs as the disappearing of the CN and $\rm C_{2}$ bandheads at
$1\lapprox \lambda/{\rm \mu m} \lapprox 1.5$ typical of C-stars in
favour of the ${\rm H_{2}O}$ molecules around $1.6~{\rm \mu m}$.
\begin{figure*}
 \psfig{figure=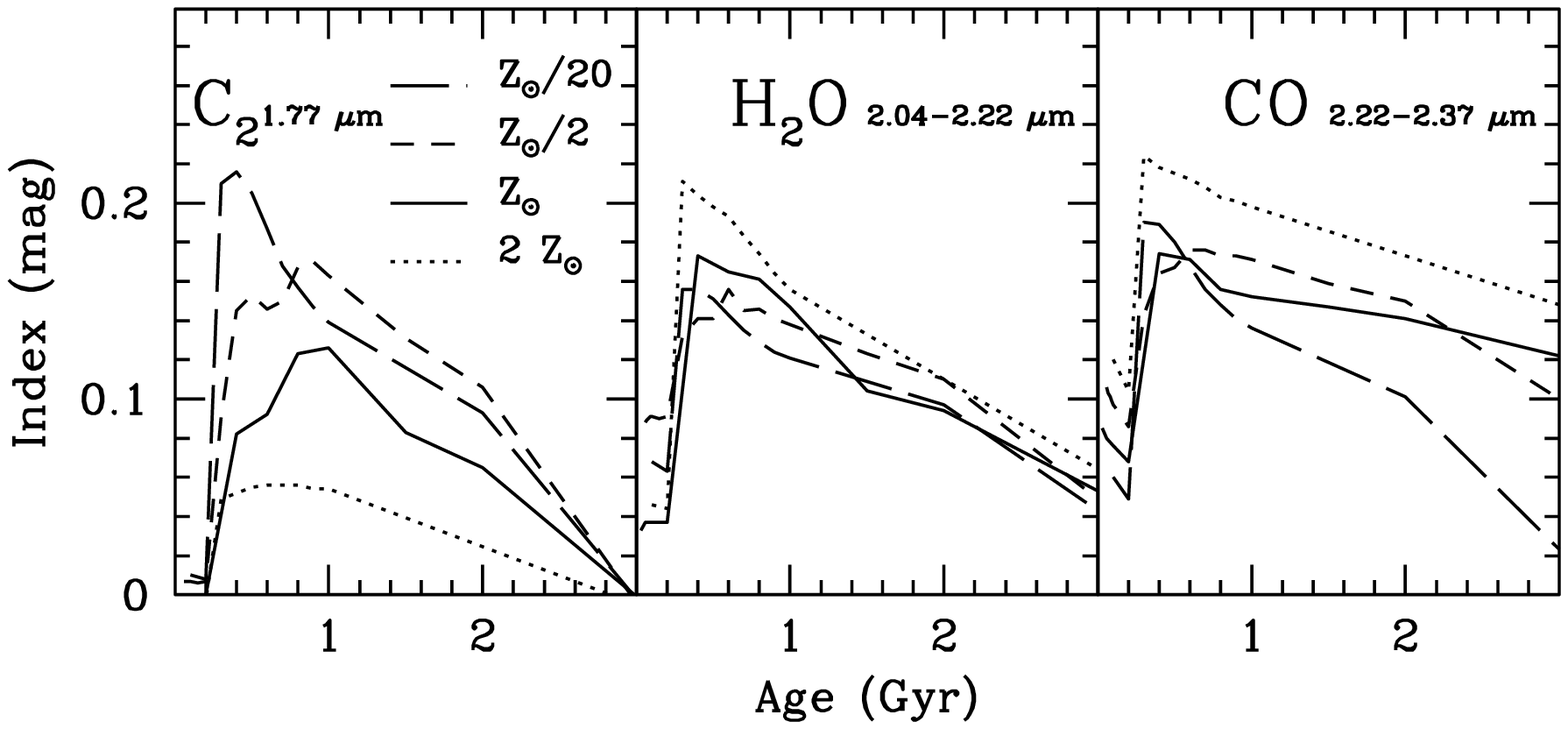,width=\linewidth}
  \caption{The IR indices $\rm C_{2}$, $\rm H_{2}O$ and CO and their
  dependence on age and metallicity. The defining wavelengths are
  indicated, indices are expressed in magnitudes.}
\label{irind}
\end{figure*}
Several line indices are defined that trace these spectral features
(e.g. Aaronson et al.~1978; Frogel et al.~1978; Alvarez et al.~2000)
and can be used in extragalactic studies. With the new spectra we are
in the position of computing the integrated indices for the SSP models
(see also \lancon~et al.~1999).

Figure~\ref{irind} shows three of these indices. The $\rm C_{2}$ index
(Alvarez et al.~2000) measures the strength of the bandhead at
$1.77~\mu m$ (Ballik \& Ramsay~1963) and is defined as the ratio
between the fluxes in the regions $1.768-1.782~\mu m$ and
$1.752-1.762~\mu m$. The water vapour absorption band index $\rm
H_{2}O$ measures the ratio of the flux densities at $2.04~\mu m$ and
$2.22~\mu m$, based on the HST/NICMOS filters F204M and F222M, or at
$1.71~\mu m$ and $1.80~\mu m$, based on the HST/NICMOS filters F171M
and F180M, and the CO index the flux densities at $2.37~\mu m$ and
$2.22~\mu m$, based on HST/NICMOS filters F237M and F222M. The indices
are defined in magnitudes and normalized to Vega\footnote{Using a
spectrum of Vega taken with HST and kindly provided by R. Bender, the
zeropoints are: $\rm C_{2}=0.038$; $\rm H_{2}O_{1.71~\rm \mu
m}=0.160$; $\rm H_{2}O_{2~\rm \mu m}=-0.360$; $\rm CO=0.263$, that
have to be subtracted to the indices}.

The $\rm C_{2}$ index is a strong function not only of the age,
peaking at the time of the AGB phase transition, but especially of the
chemical composition, its value decreasing with increasing
metallicity, following the decrease of the fuel in C-stars. The water
vapour is rather insensitive to the chemical composition, hence is a
good age indicator for ages between 0.4 and 2 Gyr. The CO index
behaves similarly to the $\rm H_{2}O$.
\begin{figure*}
 \psfig{figure=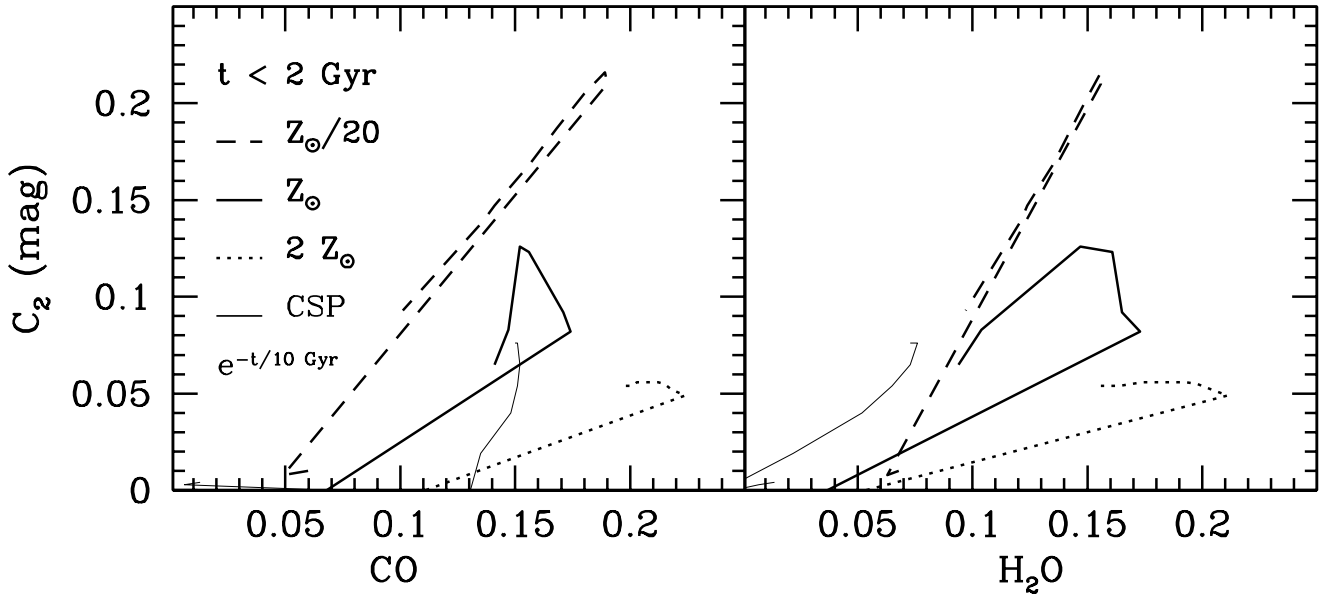,width=\linewidth}
  \caption{Diagnostic index-index diagrams $\rm C_{2}$ vs. CO
  (left-hand panel) and $\rm C_{2}$ vs. $\rm H_{2}O$ (right-hand
  panel), for stellar populations younger than 2 Gyr. The thick lines
  are SSPs with various metallicities, the thin line shows the effect
  of an extended exponentially-declining star formation history, $\rm SFR
  \propto e^{-t/\tau}$, with $\tau=10~\rm Gyr$.}
\label{irindirind}
\end{figure*}

The combination of indices that works best to unveil the presence of
$\sim~1~\rm Gyr$ stellar populations and their chemical composition is
$\rm C_{2}$ together with $\rm H_{2}O$, as shown in
Figure~\ref{irindirind}. In fact this diagram not only splits very
nicely the metallicity effects, but it separates istantaneous bursts
(SSPs, thick lines) from stellar populations being formed with
extended star formation, as shown by the thin line for an
exponentially-declining star formation history, $\rm SFR \propto
e^{-t/\tau}$ with $\tau=10~\rm Gyr$.

\subsubsection{Comparison with literature and data}
\label{compsedTP-AGB}
\begin{figure}
 \psfig{figure=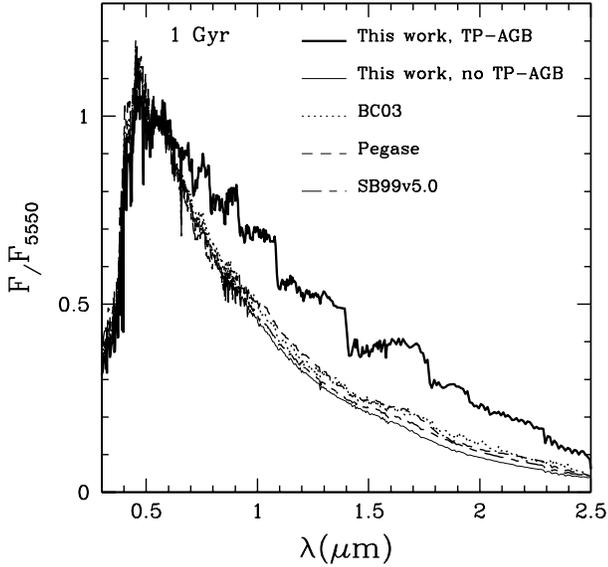,width=\linewidth} 
\caption{Comparison of the 1 Gyr, solar-metallicity SED of this paper
(thick line) with other models from the literature, from Bruzual \&
Charlot~(2003, dotted line), from the Pegase code (PEGASE.2, dashed
line) and from the latest version of Starburst99 (V\'azques \&
Leitherer 2005, long-dashed-short-dashed line). Our SED without TP-AGB
phase is shown as a thin solid line}
\label{sedlit}
\end{figure}
Figure~\ref{sedlit} shows our SED of a 1 Gyr, $Z=Z_{\odot}$ SSP (thick
line) and similar models from Bruzual \& Charlot~(2003, hereafter
BC03), from the Pegase code (Fioc \& Rocca-Volmerange~1997, 
version 2 as available at http://www.iap.fr/pegase, hereafter PEGASE.2
models) and from the latest version of Starburst99 (V\'azques \&
Leitherer 2005, long-dashed-short-dashed line). The BC03, PEGASE.2 and
Starburst99 are very similar one to each other which is due to their
use of the Padova tracks, and are more consistent with the version of
our models not including the TP-AGB phase than with that in which the
TP-AGB is accounted for. This result is not easy to understand. In the
BC03 models the inclusion of C-stars is accounted for by means of
unpublished theoretical spectra and an unspecified
temperature/luminosity relation, therefore it is hard to make a
meaningful comparison with our models. In the PEGASE.2 models and in the
latest version of the Starburst99 models, the TP-AGB phase is included
by adopting theoretical prescriptions for luminosities and lifetimes
(Gronewegen \& de Jong~1993 and Vassiliadis \& Wood 1993,
respectively), but the key information concerning the TP-AGB fuel
consumption as a function of mass, its comparison with the Frogel,
Mould \& Blanco data, the inclusion of Carbon stars and which spectra
are assigned to TP-AGB stars are not specified.

We now compare the model SEDs with Magellanic Clouds GCs, that are the
ideal templates in the age range relevant to the TP-AGB. In M98 we
used the data available at the time, namely $U,B,V,K$. Here we
complete the wavelength coverage by adding data in the $R$,
$I$-Cousins bands from new observations of a sample of MC GCs in the
relevant age range (Goudfrooij et al.~2005, {\it in preparation}). For
this sample we can perform the SED fittings.  These are shown in
Figure~\ref{lmcsed}, in which nine out of the 20 objects of the
Goudfrooij et al.~(2005) sample are considered, according to the
availability of all bands and also to reasons of space, since some
objects display basically the same spectral energy distributions. The
selected GCs span the interesting range in SWB (Searle, Wilkinson \&
Bagnuolo~1980) types from 3 to 5.5. The SWB scheme is a ranking of the
MC GCs in terms of age/metallicity, the selected types referring to
the age range $0.1~\lapprox~t/{\rm Gyr}\lapprox~2~\rm Gyr$ that is
relevant to the model comparison. GCs with greater types are older and
more metal-poor. The type 7, for example, corresponds to Milky
Way-like objects, with ages $\sim 13~\rm Gyr$ and $\zh\sim-2$. As
discussed by Frogel, Mould \& Blanco~(1990) the exact ranking of some
individual clusters might be not so meaningful, but the range from 4
to 5.5 corresponds to objects with the largest numbers of AGB stars,
that are the only ones in which Carbon stars are detected (see Fig.~3
in M98). Therefore the GCs in this SWB range span average ages between
$\sim~0.3$ and $\sim 2$ Gyr (cf. Table~3 in Frogel, Mould \&
Blanco~1990). Indeed, ages determined by various methods agree
generally well with the SWB ranking. For example, in order to fit
NGC~1987 we use an SSP with the age and the metallicity as determined
in the literature (from Girardi et al.~1995 and Ferraro et al.~1995),
and the result is very good. However we emphazise that with the
exercise of Figure~\ref{lmcsed} we do not aim at deriving precise ages
for the objects, but rather at illustrating the effect of the TP-AGB
phase. Figure~\ref{lmcsed} shows that their SEDs can be fit only with
a proper inclusion of the TP-AGB energy contributions and the spectra
of C- and O- stars.

The SSPs have metallicities either half-solar or $1/20~\zsun$
(i.e. $\zh=-0.33$ and $\zh=-1.35$), with thick lines showing our
models, and dotted and dashed thin lines those by BC03 and
PEGASE.2, respectively. For the latter the lower metallicity is
$\zh\sim-0.7$. For our models we show also the broad-band fluxes
(empty circles). Each panel presents an individual GC SED (filled
points). Starting from the top left, object NGC~265 with ${\rm SWB}=3$
is a pre-AGB GC whose SED does not yet display evidences of TP-AGB
stars redwards the $R$-band. The situation changes completely when
later SWB types are considered and the typical features of the cool
TP-AGB stars appear in the near-IR spectrum. Their {\it complete}
spectral energy distributions are well fit by our models.

\begin{figure*}
 \psfig{figure=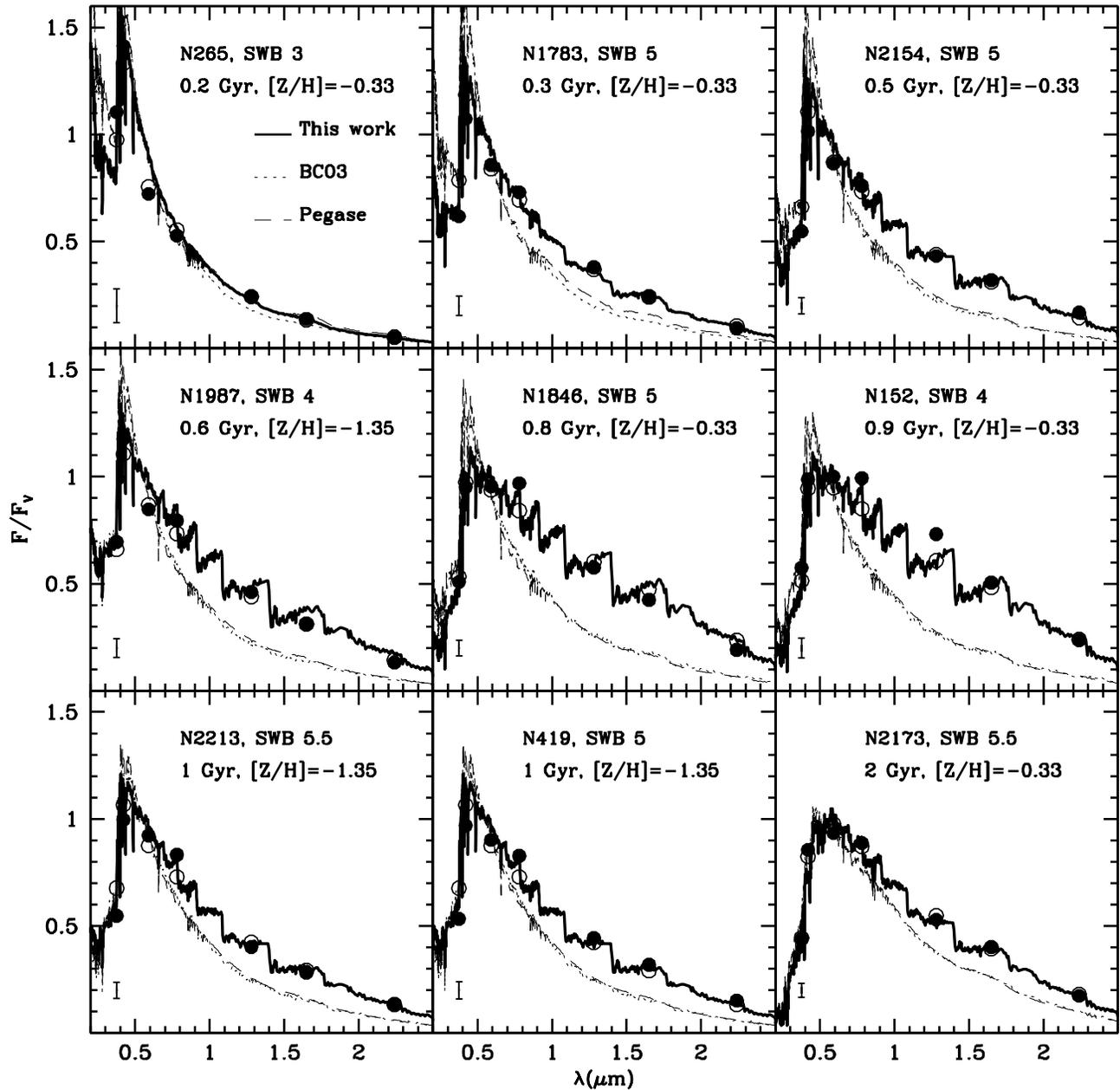,width=\linewidth} 
\caption{SED fitting of MC GCs data (filled points). In each panel the
cluster ID and SWB type (the latter from Frogel, Mould \& Blanco~1990)
are indicated. Sources of data are: van den Bergh~1981 for $U,B,V$;
Persson et al.~1983 for $J,H,K$; Goudfrooij et al.~2005 for
$V,R,I$. The reddening $E(B-V)$ is taken either from Persson et
al.~(1983) or from Schlegel et al.~(1998). The largest error (on the
flux in the $B$-band) is shown.  Lines show ours, the BC03  and
the PEGASE.2 models (solid thick, and dotted and dashed thin lines,
respectively) and the ages/metallicity of the fits. For our models
also the broad-band fluxes are given (empty circles).}
\label{lmcsed}
\end{figure*}
The models by BC03 and PEGASE.2 do not reproduce the observed SEDs
with the same SSP parameters, displaying substantially less flux
redward the $R$-band. The same comparison holds for the Starburst99
models (not shown, see Figure~\ref{sedlit}). The obvious
conclusion is that the recipes for the TP-AGB adopted in those models
are not adequate at describing real stellar populations with TP-AGB
stars. Further discussions on these models is found in the next
section.  There is no combination of age and metallicity that allows
to fit the high fluxes both blueward and redward the $V$-band unless
the TP-AGB phase is included as shown by the thick solid lines. As we
will see in Section~\ref{highztpagb}, in case of galaxies a composite
stellar population in which a late burst is superimposed to an old
component can fulfill the need for high blue and near-IR fluxes
(relative to $V$). This option is clearly not viable in case of GCs.

In the next section we will show the comparison with more objects by
means of broadband colours.

\subsection{Broadband colours}
\label{colors}
Broadband colors are computed for several filter systems
(e.g. Johnson-Cousins, SLOAN, HST). In the following subsections we
will compare them with Magellanic Clouds GCs and Milky Way Halo and
Bulge GCs, which allows the check of young and intermediate age models
with slightly subsolar metallicities, and old ages for a wide range of
metallicities, respectively. Comparisons will be also made with models
from the literature.
\subsubsection{TP-AGB-dominated $\sim 1~\rm Gyr$ models}
\label{colorsTP-AGB}
\begin{figure*}
 \psfig{figure=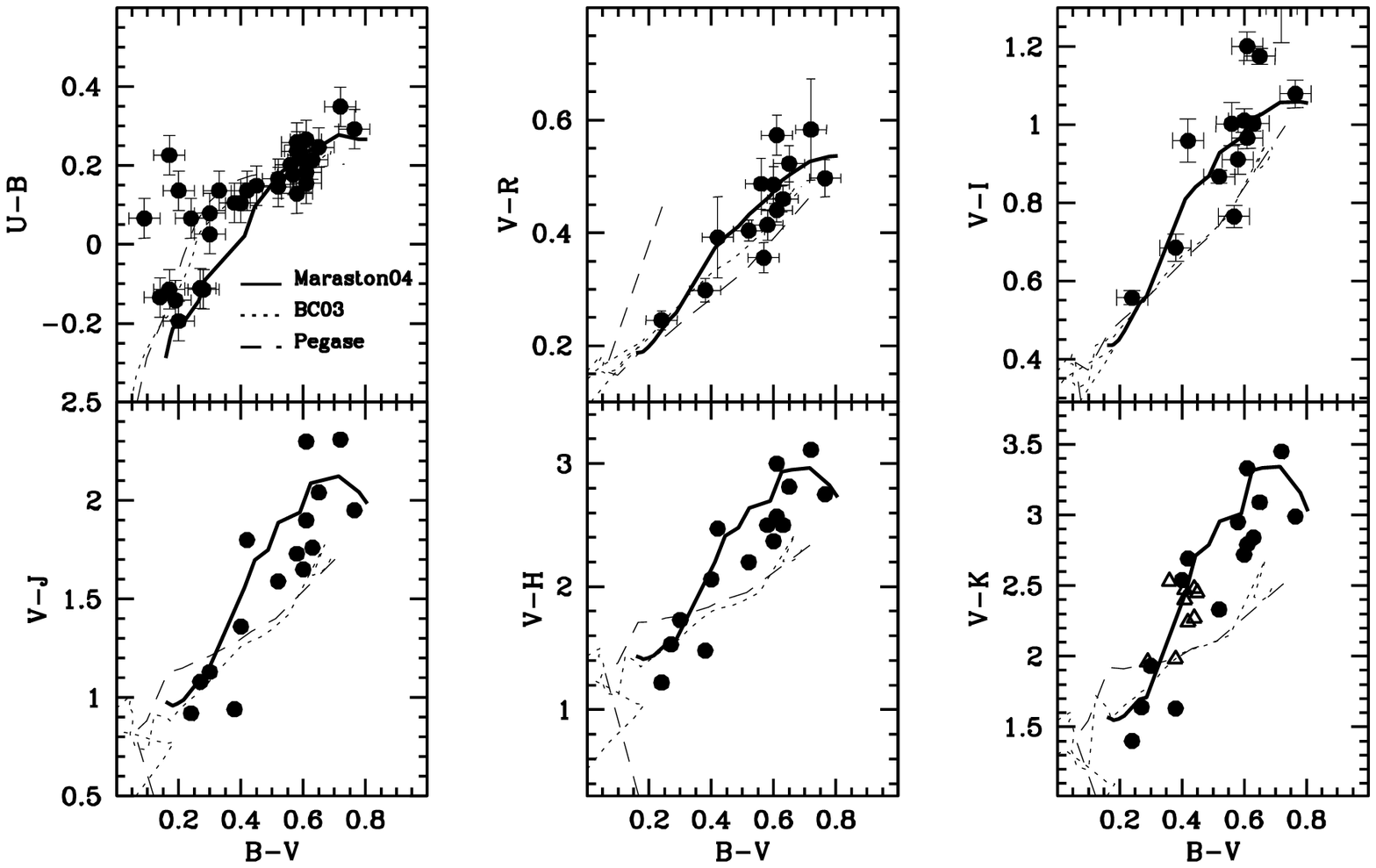,width=\linewidth}
  \caption{Calibration of TP-AGB dominated synthetic broadband colours
  with data of MC GCs (filled circles). MC data are selected to belong
  to SWB types between 3 to 6, appropriate to the age range and
  the metallicity of the models (sources of data as in
  Figure~\ref{lmcsed}). Open triangles show the star clusters of the
  merger remnant galaxy NGC 7252 ($V$ photometry from Miller et
  al.~1997, $K$ photometry from Maraston et al.~2001). Models have
  ages lower than 2 Gyr and metallicities $0.5~\zsun$, with solid,
  dotted and dashed lines referring to this work, BC03 and PEGASE.2. }
\label{compagb}
\end{figure*}
In Figure~\ref{compagb} we perform a comparison similar to that shown
in Figure~\ref{lmcsed} by means of broadband colours which allows us
to use a larger data sample. We also include the data of the GCs in
the merger remnant galaxy NGC 7252 (Maraston et al.~2001, open
triangles). The models are selected to have ages smaller than 2 Gyr
and half-solar metallicity.

The AGB phase transition among the clusters appears as an enhacement
of the IR luminosity with respect to the optical one, that increases
with increasing wavelength, with e.g., the $V-K$ colour reaching
values larger than 3. All bands redwards $R$ are affected. The models
of this paper provide a good description of TP-AGB-dominated stellar
populations in a large wavelength range. As already shown in Maraston
et al.~(2001), the models reproduce very well also the integrated
colours of the young GCs of the merger remnant galaxy NGC~7252.

The other models considered here, namely those by BC03 and
PEGASE.2\footnote{For this comparison we focus on those models for which
the TP-AGB is stated to be included. The Worthey (1994), the Vazdekis
et al. and the Starburst99 (Leitherer et al.~1999) models do not treat
the TP-AGB phase. Mouhcine \& \lancon~(2002) include the TP-AGB phase
in such a way as to reproduce the $B-V$ vs. $V-K$ synthetic diagram
published by Maraston et al.~(2001). However the TP-AGB bolometric
contribution in their models is up to 20 per cent (Mouhcine et
al.~2002), smaller than the observed $\sim 40$ per cent one (Frogel,
Mould \& Blanco~1990) reproduced by our models
(Figure~\ref{contrpha}).} do not exhibit the ``jump'' in the near-IR
colours displayed by the MC GCs, remaining systematically bluer than
the data and evolving slowly towards the red colours produced by the
rise of the RGB phase at $t\gapprox~1~\rm Gyr$. This pattern is
equivalent to the SED comparison discussed in the previous section,
again suggesting that the TP-AGB phase is not fully accounted for in
these models.

BC03 argue that stochastical fluctuations in the number of TP-AGB
stars suffice to explain the full range of observed $V-K$ colours of
MC GCs. These effects are mimicked by means of a stochastic IMF in a
$\sim~2\cdot 10^{4}\msun$ model star cluster.

From their Figure~8 one sees that the simulations match very well the
colours of the youngest MC GCs, whose near-IR light is dominated by
red supergiants with lifetimes $\sim~10^6~\rm yr$. The TP-AGB phase
instead lasts ten times longer and is therefore less affected by
stochastical fluctuations. As a consequence the simulations are less
successful in covering the observed colours of the TP-AGB-dominated MC
star clusters. Moreover, the probability distributions of the points
in their simulations is squewed towards bluer colors. This implies
that it would be highly unlikely to observe a GC on the red side of
the models if stochastical fluctuations were dominating the
distribution of the data. Instead the data scatter exactly to the red
side of the models.

A further important point is that the importance of stochastical
fluctuations depends dramatically on the mass of the globular
cluster. While the effect is relatively large for masses of the order
$10^4\msun$ (considered in the BC03 simulations), it is significantly
smaller at $10^5\msun$ and completely negligible around
$10^6\msun$. Hence, the stochastical fluctuations, based on a
$10^4\msun$ cluster, are not appropriate to describe the colours of
the star clusters of the merger-remnant galaxy NGC~7252 shown as open
triangles in Figure~\ref{compagb}. Their luminous masses are estimated
to be at least $10^6~\msun$ ~(Schweizer \& Seitzer~1998) ranging up to
even $\sim 10^8\msun$ as confirmed dynamically for the most luminous
object (Maraston et al.~2004). The expected stochastical fluctuations
for such very massive objects are of the order of a few percent
(Maraston et al.~2001) and their colours are not explained by the BC03
models without fluctuations (Figure~\ref{compagb}). Instead, their
colours are nicely explained by a TP-AGB phase like in the MC clusters
(Maraston et al.~2001 and Figure~\ref{compagb}), a conclusion that is
supported by the direct observation of Carbon stars in their spectra
(Mouhcine et al.~2002 with SOFI observations).

To conclude, while we fully agree that stochastical fluctuations in
the number of stars along {\it short} evolutionary
phases\footnote{e.g. the supergiant phase in stellar populations
younger than 0.1 Gyr} scatter the near-IR colours of small-mass star
clusters, the account of these effects should be considered on top of
the stellar evolutionary phases. For example it seems more likely to
us that the lack of the AGB phase transition in the BC models comes
from the fact that the TP-AGB bolometric contribution never exceeds
$\sim~10$ per cent (Figure~4 in Charlot \& Bruzual~1991). This is at
variance with the measured bolometric contribution of the TP-AGB phase
($\sim~40$ per cent, Figure~3 in M98) which was evaluated by taking
stochastical fluctuations into account.

In the PEGASE.2 models the details of the inclusion of the TP-AGB phase
are not specified. However from Figure~2 in Fioc \&
Rocca-Volmerange~(1997) one sees that a jump in $V-K$ of
$\sim~0.6$ mag (up to $V-K\sim 2.2$) occurs at $t \sim~0.1$ Gyr, after
which the evolution of this colour is nearly constant. The AGB
phase-transition as exhibited by the MC star clusters does not take
place in their models.
\subsubsection{Age and metallicity relations of old models}
\begin{figure*}
 \psfig{figure=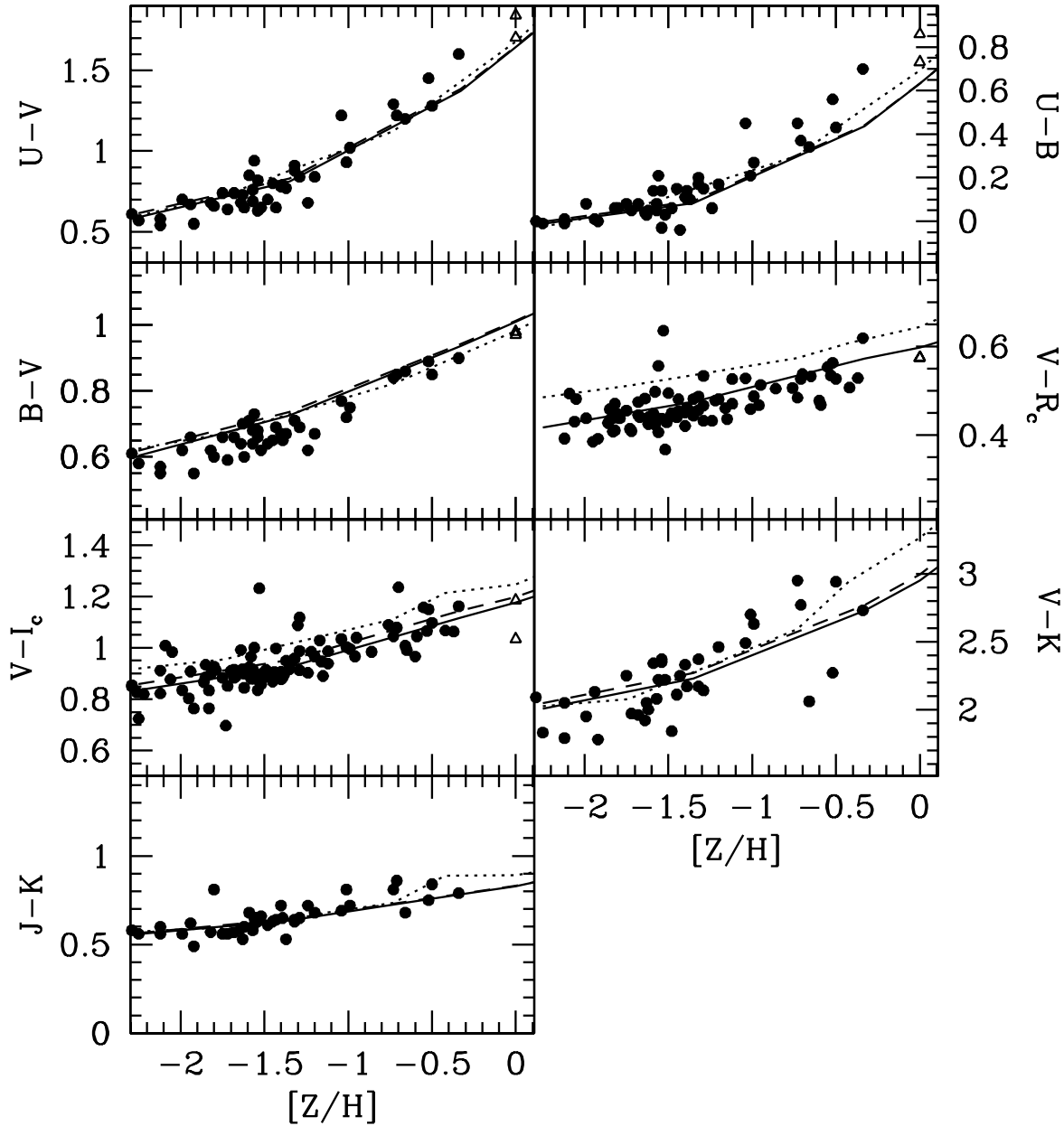,width=\linewidth}
  \caption{Comparisons of synthetic broadband colors with data of
  Milky Way GCs (as compiled in the Harris - 1996 and revisions -
  catalogue). The data of the $\sim~\zsun$ Bulge GCs NGC 6528 and NGC
  6553 are shown as open triangles. Reddening has been subtracted as
  in Barmby \& Huchra ~(2000). Metallicities are on the Zinn \&
  West~(1984) scale. Solid lines show SSP models of age 13 Gyr, Kroupa
  IMF, and blue/intermediate HB morphology at $\zh\lapprox~-1$. The
  same models with a Salpeter IMF are shown as dashed lines. The
  dotted lines are the BC03 models for the same age.}
\label{compgcsmw}
\end{figure*}
Milky Way GCs spanning a wide range of metallicities at nearly the
same age are ideal to calibrate the synthetic colours/metallicity
relations. Figure~\ref{compgcsmw} shows a comprehensive comparison of
old (13 Gyr) models with all available photometry of Milky Way
GCs. Data are plotted towards the \feh~parameter of the Zinn \&
West~(1984) scale, that we show to reflect total metallicities
(Thomas, Maraston \& Bender~2003). Data for the two metal-rich GCs of
the galactic Bulge NGC 6528 and NGC 6553 are the open
triangles. Models are shown for both the Kroupa and the Salpeter IMF
(solid and dashed lines, respectively) only for completeness since we
found that the optical colors of SSPs are virtually unaffected by
plausible IMF variations (M98). The dotted line shows the BC03 models
for the same age and the Salpeter IMF.

The match between models and data is very good, except for the $B-V$
colour, that appears to be systematically redder in the models (of
0.05 mag). We have checked that this feature is common to all SSP
models considered in the paper, namely the BC03, PEGASE.2, Worthey
~(1994) and Vazdekis et al.~(1996), and was discussed by
Worthey~(1994) and Charlot, Worthey \& Bressan~(1996). We now
investigate the origin of the offset.

The offset cannot be attributed to HB morphologies. Moreover, the blue
HB morphology\footnote{as derived from the HBR parameter listed in the
Harris~(1996)} of most GCs with metallicities below $-1.3$ is already
accounted for in our models at 13 Gyr and $\zh=-2.25$
(cf. Figure~\ref{hbmor}). At 13 Gyr and $\zh=-1.35$ the HB morphology
is intermediate (cf. Figure~\ref{hbmor}). A purely blue HB, with
$\teff$ up to $10^4$ K would make the $B-V$ bluer by 0.03 mag. But the
GCs with metallicities $\gapprox -1.$ have a red HB and still their
$B-V$ is bluer than the models. The latter data can be better match
with a younger age, e.g. 8 Gyr, which is perhaps not excluded
(e.g. Rosenberg et al.~1999). However this younger age worsens the fit
to $U-V$ and $U-B$.

A more likely explanation is a defect in the colour-temperature
transformations. This issue is extensively addressed by Vandenberg \&
Clem~(2003), who provide a set of transformations to cure the $B-V$
problem, which is attributed to deficiencies of the model
atmospheres. We have found that the $B-V$ colors in their
transformations are very close to those of the BaSel library for the
turnoff temperatures of the 13 Gyr models ($5800\lapprox \teff/{\rm
K}\lapprox~6700$, $\rm log g \sim 4.5$), but are {\it bluer} for the
RGB base ($4900\lapprox \teff/{\rm K}\lapprox~5300$, $\rm log g \sim
3$), by 0.1 mag at $\zh\sim -2.25; -1.35$.  Since the RGB contributes
$\sim~40$ per cent to the $V$-luminosity at $\zh=-1.35$ and below
(cf. Figure~\ref{contrpha}) the use of the Vandenberg \& Clem~(2003)
transformations would be able to reduce significantly the discrepancy
between data and models. It is intriguing however that the Vazdekis et
al. models that adopt an empirical temperature-colour relation instead
of the theoretical one suffer from the same problem. Also, Bruzual \&
Charlot~(2003) compare the integrated $B-V$ of solar metallicity SSPs
based on the BaSel and the Pickles empirical library, and found
negligible differences. Though their exercise refers to solar
metallicity, while our comparison focuses on sub-solar chemical
compositions, it has to be expected that if a fundamental problem in
the treatment of the opacity would be responsible for the colour
offset, this would be even more serious at higher metallicities.

We are left with one alternative explanation, that is an offset in the
photometric bands, due to the broadband filters. Indeed, two different
B~filters are used, the so-called B2~and B3 (Buser), in order to
compute $U-B$ and $B-V$, respectively. These are commonly defined as
``standard Johnson B''. The colour $U-V$ is then evaluated by adding
$U-B$ and $B-V$\footnote{We acknowledge a discussion with R. Buser who
confirmed this fact to us.}. It is hard to trace back which exact
filter was used for the data, therefore we vary the B-filter in the
models. The result is that if we use the B2 filter to compute the
$B-V$ colour, the latter gets {\it bluer} by 0.03 mag., again in the
direction of improving the comparison, while leaving the $U-V$ colour
almost unchanged.

To conclude, either a defect of the model atmospheres or a filter
mismatch or a combination of both are the most plausible sources of
the discrepancy between data and models in $B-V$. However since at
redshift $>0$ one needs the whole spectral energy distribution, the
use of spectral libraries like the BaSel one is unavoidable. Instead,
in comparing the models with observed colours, it is safer to take the
offset of 0.05 mag. into account.

Back to Figure~\ref{compgcsmw} we see that the BC03 models behave very
similarly to ours, with the exception of the $V-R$ and $V-I$ colours,
that in the BC03 models appear to be too red when compared to our
models and to the data. Reducing the age does not improve the
comparison. We cannot explain the reasons for this offset.
\subsection{Mass-to-light ratios}
\label{ml}
The stellar mass-to-light ratios $M^*/L$ are computed as in M98, by
taking the stellar mass losses into account. The relation between
living stars and remnants is from Renzini \& Ciotti~(1993), where the
remnant mass as a function of the initial mass is given for three
types of stellar remnants, i.e. white dwarfs, neutron stars and black
holes. These relations are used for every metallicity, the dependence
on the chemical composition entering through the turnoff mass.
\begin{figure}
 \psfig{figure=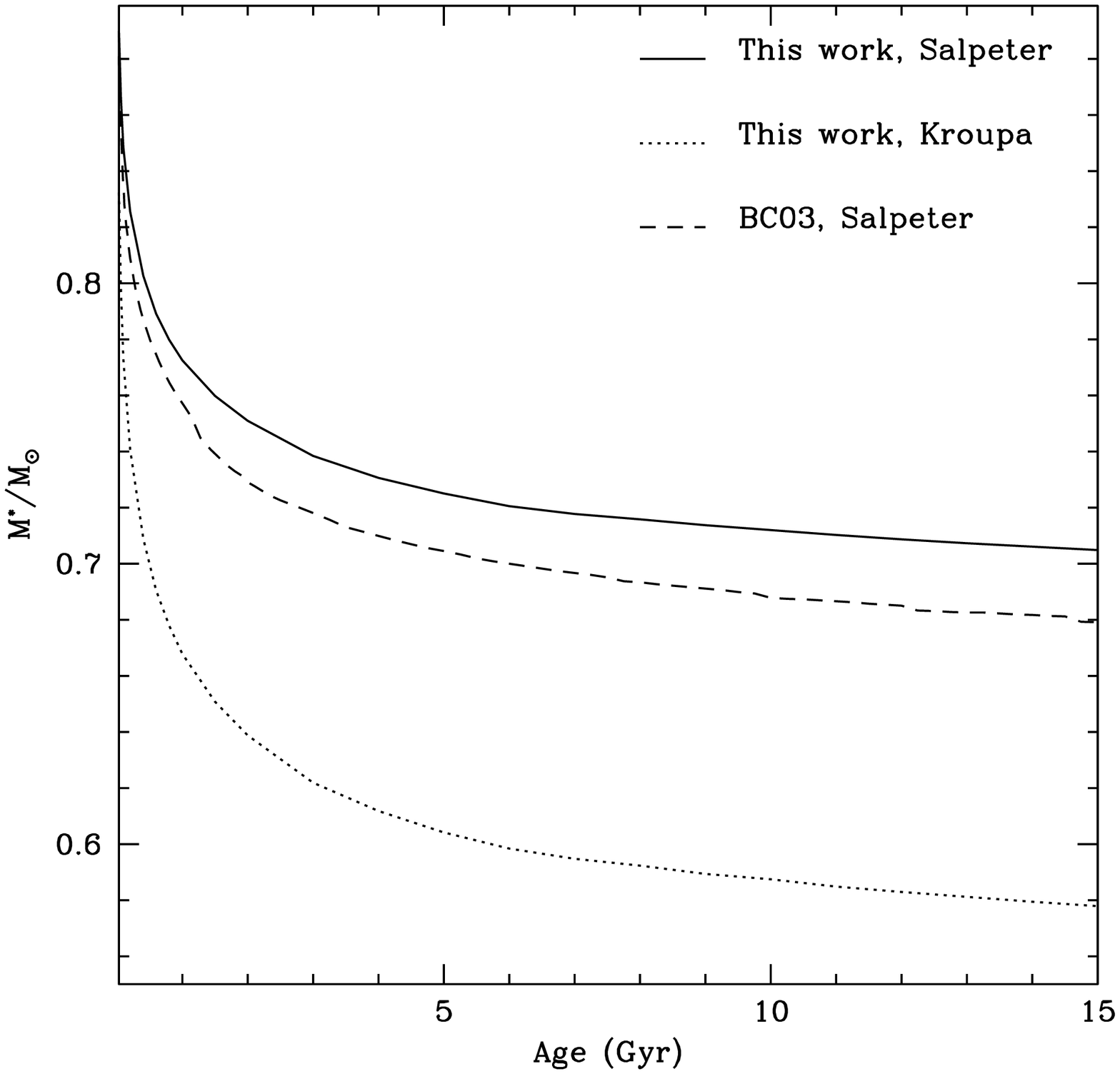,width=\linewidth}
  \caption{The total stellar mass (normalized to $1~M_{\odot}$) of an
  evolving stellar populations in which the stellar mass losses are
  taken into account.}
\label{mstar}
\end{figure}
Figure~\ref{mstar} shows the stellar mass of an evolving stellar
population with initial mass of $1~\msun$, solar metallicity and
Salpeter or Kroupa IMFs (solid and dotted line, respectively). In case
of a Salpeter IMF the stellar mass decreases by $\sim 30$ per cent in
15 Gyr, $\sim 23$ per cent being lost in the first Gyr. The Kroupa IMF
with less power in low-mass stars has $\sim$ 16~per cent~less mass,
and correspondingly lower $M^*/L$'s. The stellar mass of the BC03
models is smaller than that computed in our models.
\begin{figure}
 \psfig{figure=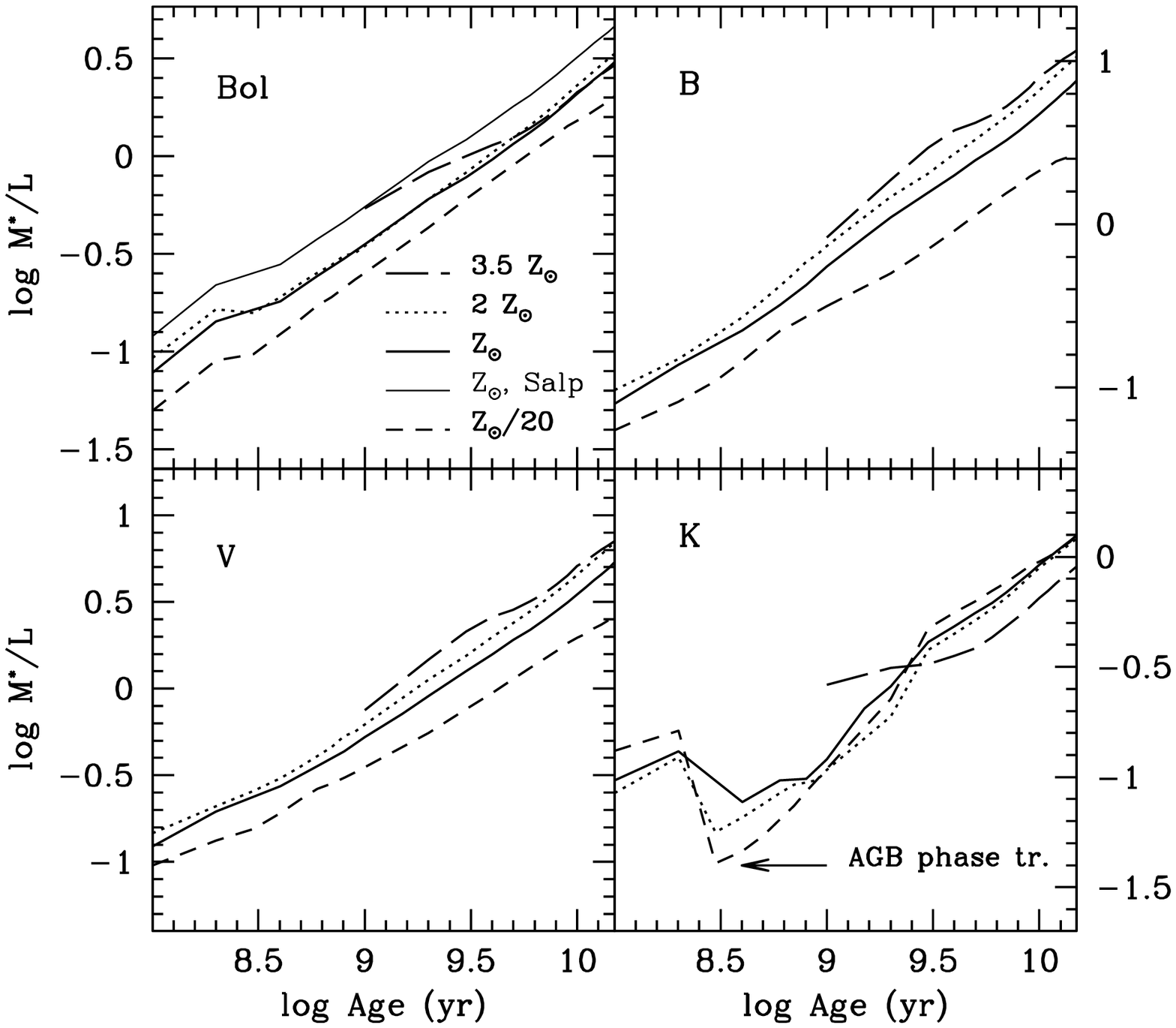,width=\linewidth}
  \caption{The dependence of the $M^{*}/L$ (in solar units) on age and
  metallicity. Models refer to the Kroupa IMF, for the bolometric the
  solar-metallicity models for a Salpeter IMF are shown as thin solid
  line. Note that the 3.5 $\zsun$ SSP is computed with the Padova
  tracks and does not contain the TP-AGB phase.}
\label{mlz}
\end{figure}

The basic dependences of the $M^*/L$'s on the SSP parameters, ages,
metallicities and the IMF are shown in Figure~\ref{mlz}. The $M^*/L$'s
increase with age because the luminosities decrease, an effect
independent of the photometric band. An important exception is the
occurrence of the AGB phase transition during which the near-IR
luminosity increases with age and therefore the $M^*/L_{\rm
near-IR}$ {\it decreases}, by a factor 3-5 with respect to models not
including the TP-AGB phase.

Metallicity effects are strongly dependent on $\lambda$. The $M^*/L$'s
generally increase with increasing metallicity because the
luminosities decrease (which is explained by lower MS and PMS
luminosities with increasing metal content,
Section~\ref{energetics}). However in the near-IR, a transition occurs
at $\lambda\gapprox$ I-band and the $M^*/L$'s become {\it independent}
of metallicity at ages $\gapprox~2$ Gyr. The reason for this is that a
high $Z$ SSP produces less light than one with a lower amount of
metals, but most of the energy is emitted preferentially at longer
wavelengths because of the cooler stellar temperatures. Note however
that the effect holds until the metallicity is not too large. The very
metal-rich SSP of our set ($Z=3.5~\zsun$)~has $M^*/L_{\rm near-IR}$
smaller than the others SSPs.

The effect of the IMF was amply discussed in M98. The main result was
that the $M^*/L$ of SSPs is minimal for Salpeter and increases for
both a flattening or a steepening of the IMF. The reason is that a
larger amount of mass in massive remnants or in living stars is
present. This interesting point stems from considering the evolution
of the stellar mass. The models in Figure~\ref{mlz} refer to a Kroupa
IMF, the effect of having a Salpeter IMF is shown only in the
bolometric (solid thin line), with the $M^*/L$ being larger by a
factor $\sim 1.5$. A similar scaling factor is found at the various
$\lambda's$, almost independent of metallicity.
\begin{figure}
 \psfig{figure=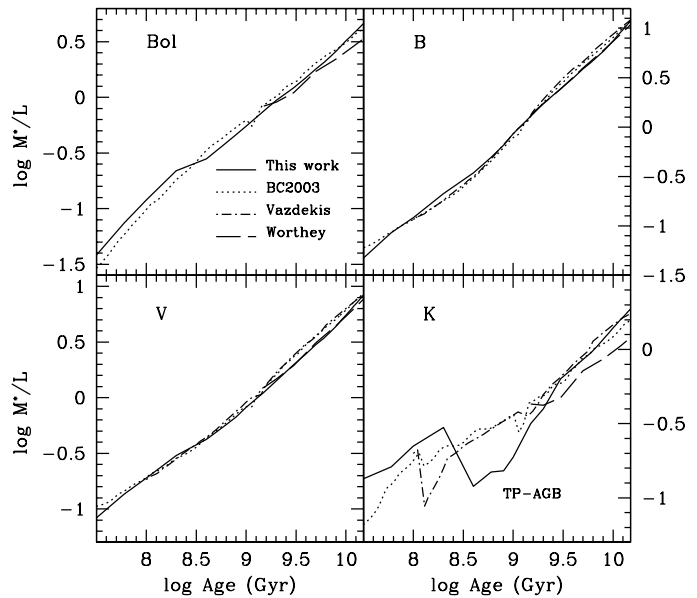,width=\linewidth}
  \caption{Comparison between model $M/L$ in various bands from
  different authors (see Footnote 12). All SSPs have solar metallicity
  and Salpeter IMF. }
\label{mlcomp}
\end{figure}

Finally our models are compared with others in the literature in
Figure~\ref{mlcomp}\footnote{During the referee process of this paper
A. Vazdekis provided us with a new version of his model $M/L$ in which
the values have been revised. This explains why Figure 24 in its final
version differs from that of the astro-ph submitted paper}. The models
by BC03 have a bolometric $M/L$ that is larger than ours around ages
of a few Gyr, because the luminosity of the Padova tracks is smaller
due to the delayed development of the RGB phase (Figure~7). The effect
is partly counterbalanced by a slightly smaller stellar mass
(Figure~\ref{mstar}). The largest differences are found in the
near-IR. The $M/L_{\rm K}$ of the BC03 models is higher than ours
around 0.8-1 Gyr due to the lower contribution of the TP-AGB in their
models. At old ages it is instead smaller, due to the cooler RGBs of
the Padova tracks (cf. Figure~9). The models by Worthey~(1994) have
smaller $M/L_{\rm bol}$ and $M/L_{\rm K}$, due to larger bolometric
and near-IR luminosities. This is possibly due to a larger number of
upper RGB stars with respect to that predicted by the Padova
isochrones (Charlot, Worthey \& Bressan~1996). Note also that the
Worthey~(1994) $M/L$'s do not take the stellar mass losses into
account. Finally, the models by Vazdekis et al.~(1996) behave in
general similarly to the BC03 ones, which is due to the use of the
Padova tracks in both models. Some differences appear in the $K$-band,
however. At $t\sim 10^{8}$ yr a dip in the Vazdekis models is present
that is not found in the BC03 models, therefore it can hardly be
connected to the Padova tracks, unless the two models adopt a
different releases.
\subsection{Lick indices}
The Lick indices of SSP models computed by means of the Lick fitting
functions (Worthey et al.~1994) have solar-scaled element abundance
ratios at high-metallicities~$\zh\gapprox 0.5~Z_{\odot}$ (Worthey et
al.~1994; Maraston et al.~2003). Therefore they are not adequate to
model stellar systems with high-metallicities and enhanced $\alpha$/Fe
abundance ratios, like Bulge globular clusters, elliptical galaxies
and bulges of spirals (Maraston et al.~2003). At low metallicities,
these models trace element-abundance ratios that are proper to the
stars used to compute the fitting functions, which is a mix of
solar-scaled and enhanced ratios (Maraston et al.~2003; Thomas,
Maraston \& Bender~2003). Finally, the models do not contain the
explicit dependence on the relative elemental proportions, which is
instead a powerful tool to contrain the star formation history of
stellar systems (Matteucci~1994; Thomas, Greggio \& Bender~1999).

To circumvent the limitations quoted above, we have computed
new-generation stellar population models of Lick indices that include
the dependence of the element ratios. The models allow several non
standard elemental mixtures, and are checked to reproduce the Lick
indices of Milky Way Halo and Bulge globular clusters for their proper
element ratios. The models of the classical Lick indices are in
Thomas, Maraston \& Bender~(2003) and those of the high-order Balmer
lines in Thomas, Maraston \& Korn (2004).
\subsection{The D-4000}
The D-4000 break is computed adopting the definition of
Bruzual~(1983). The main features of this index have been outlined by
several authors (among others Bruzual~1983; Barbaro \& Poggianti~1997;
Gorgas et al.~1999) and will not be repeated here.  What we add are
two interesting behaviours.

\begin{figure*}
 \psfig{figure=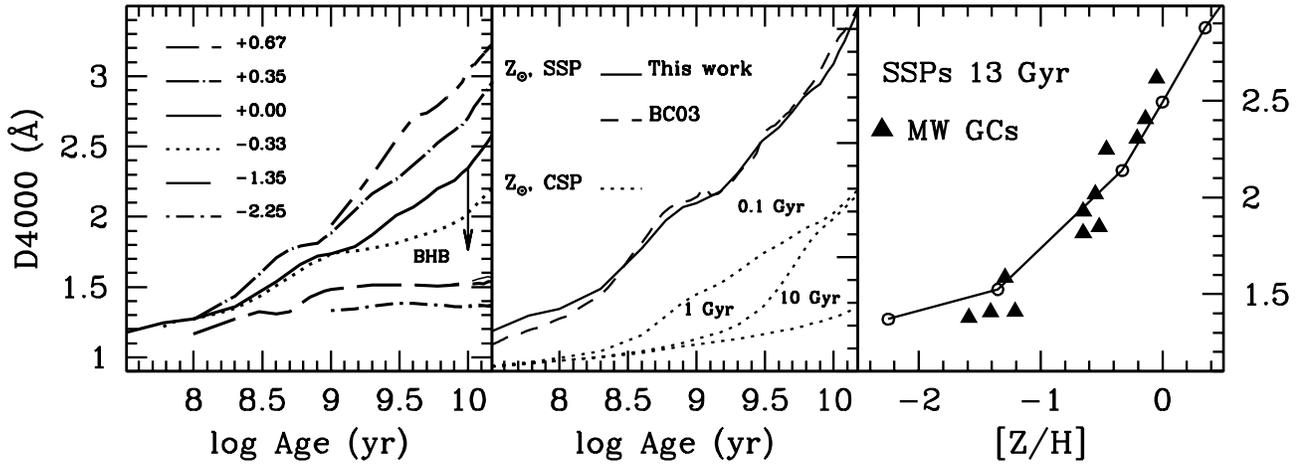,width=\linewidth}
  \caption{The D-4000 break all at once. {\it Left-hand panel}. The
  dependence on age, metallicity and Horizontal Branch. The two lowest
  metallicity SSPs are shown with both blue and red HBs (thick and
  thin lines, respectively), and a negligible impact is found. The
  arrow points to the value of the break when a blue HB is considered
  in the 10 Gyr, solar metallicity SSP. A sizable decrease of
  $\sim~0.6$ \AA~occurs as a consequence of a blue instead of a red
  HB. {\it Central panel}. The dependence of the D-4000 break on the
  input stellar tracks, shown using the BC03 models and solar
  metallicity SSPs, is negligible. The dependence on the star
  formation history is highlighted by means of Composite Stellar
  Population (CSP) models with various exponentially-decreasing star
  formations (dotted lines, the e-folding times are labelled beside
  each curve). {\it Right-hand panel} Comparison of the synthetic
  D-4000 break of SSPs with various metallicities and the constant age
  of 13 Gyr with GC data.}
\label{d4000}
\end{figure*}
Figure~\ref{d4000} summarizes the main features of the D-4000
break. Typically used as pure age indicators, the D-4000 is very much
sensitive to metallicity (Figure~\ref{d4000}, left-hand panel), and at
low metallicity is actually completely insensitive to age.
Interestingly the D-4000, unlike the Balmer lines (Maraston \&
Thomas~2000) is {\it insensitive} to the morphology of the Horizontal
Branch at low metallicity. This suggests its use as metallicity
indicator for low-metallicity extragalactic GCs, since at low
metallicity the break is also completely insensitive to age. In
contrast we find the break to be very sensitive to blue HB at high
metallicity. This is also shown in Figure~\ref{d4000}, left-hand
panel}, where the arrow indicates the value of the break of a 10 Gyr,
solar metallicity model with a BHB (see Section~\ref{bhbhighz}). The
D-4000 break is dimished by $\sim 0.6$ \AA.
The break is completely insensitive to the IMF, when modified from
Salpeter and Kroupa (Figure~\ref{d4000}, central panel), which is
obvious because the blue spectral region is not sensitive to the
percentage of stars with masses lower than $\sim~0.5~\msun$.
 
The presence of several metallic lines in the wavelength windows
defining the D-4000 break makes this index possibly dependent on the
details of the chemical abundance pattern. Such an effect is likely to
occur at metallicities larger than solar, since the solar-scaled
models and the $\afe$-enhanced GCs data agree up to solar metallicity
(Figure~\ref{d4000}, right-hand panel). This issue is investigated in
detail in a forthcoming paper.
\subsection{The Ca II triplet index}
The equivalent width of the triplet absorption line of calcium at
$8600$ \AA~(CaII) is computed adopting the index definition by Cenarro
et al.~(2001a). The index $\rm CaT^{*}$  is given by~$\rm
CaT^{*}=CaT-0.93 \cdot PaT$, therefore it is decontaminated from the
possible contribution by the adiacent Paschen lines (described by
means of the index PaT). The integrated indices of the SSPs are
computed by adopting the index calibrations with Milky Way stars
(so-called fitting functions) by Cenarro et al.~(2001b; 2002),
following a standard procedure to compute absorption line indices by
means of fitting functions. Like in the case of the Lick indices, also
the model CaT reflects the Ca abundances in Milky Way halo and disk
stars by construction.

The dependence of the CaII index on stellar parameters has been
discussed by several authors, among the first by Diaz, Diaz \&
Terlevich~(1989) till the recent papers by Moll\`a \&
Garc\`ia-Vargas~(200) and Vazdekis et al.~(2003), and by Schiavon,
Barbuy \& Bruzual~(2000), who computed the CaII index on model
atrmospheres. Briefly recalling the index is very sensitive to
gravity, being strong (weak) in cool giants (dwarfs), therefore it is
classically used to investigate the IMF. The effect of Calcium
abundance is instead still unexplored. We plan to address this issue
in a future paper.
\begin{figure*}
 \psfig{figure=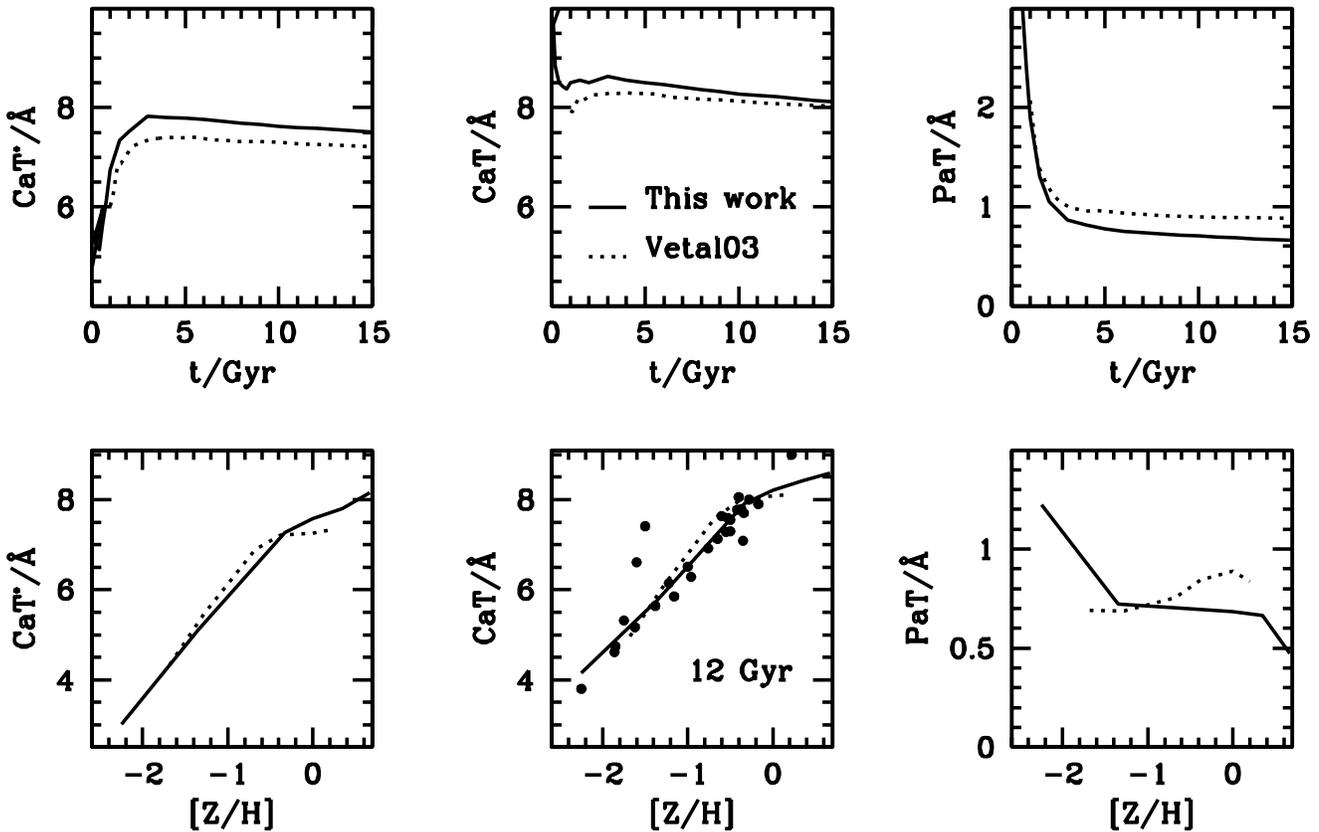,width=\linewidth}
  \caption{The dependence of the indices $\rm CaT^{*}$, CaT\ and PaT
  on age at fixed solar metallicity (upper row), and on metallicity at
  the fixed age of 12 Gyr (lower row). Solid and dotted lines refer to
  the models of this paper and of Vazdekis et al.~(2003),
  respectively, both for a Salpeter IMF. In the lower central panel
  the models are compared with Milky Way GC data (from Armandroff \&
  Zinn~1988).}
\label{cat}
\end{figure*}

The dependence of the indices on the IMF is discussed in Saglia et
al.~(2002) and Vazdekis et al.~(2003) and will not be repeated here,
where instead we focus on age and metallicity. Figure~\ref{cat} shows
the integrated indices $\rm CaT^{*}$, CaT, PaT of the SSPs as
functions of age for solar metallicity and as functions of metallicity
for the fixed age of 12 Gyr (upper and lower rows, respectively). The
indices are rather insensitive to age at $t \gapprox 2-3~\rm Gyr$,
which stems from the constancy of the giant component of a stellar
population after the RGB phase transition is completed
(cf. Figure~\ref{contrpha}). The very mild decrease of the indices at
high ages is due to the larger contribution of the low-mass
MS. Therefore the effect depends on the assumed IMF, and disappears
for IMFs with flat exponents below $0.6~\msun$. In the lower panels we
display the strong metallicity dependences of the indices, caused by
the fact that CaT measures essentially the RGB population. The
saturation of the indices at high $Z$  is most likely caused by
saturation of the CaII lines at high $Z$. Finally the central lower
panel compare the models with GCs data by Armandroff \& Zinn~(1988,
translated into the Cenarro et al. system by Saglia et al.~2002). The
models reproduce well the index-metallicity relation of the Milky Way
GCs up to a metallicity nearly solar.
\subsection{Effect of blue HBs at high metallicities}
\label{bhbhighz}
The HB at high metallicity is usually red, however the presence of
exotic blue morphologies cannot be excluded. For example, as already
mentioned in Section~\ref{HB}, two rather metal-rich Milky Way GCs
were indeed found to contain 10 per cent of blue HB stars (Rich et
al.~1997). Since in extra-galactic studies the HB morphology is not
directly accessible, it is useful to evaluate theoretically the effect
on the SSP models of a BHB at high $Z$ .
\begin{table*}
 \centering \caption{\label{BHBtab} Effects of high $Z$ BHBs on
  various SSP outputs. Numbers refer to the percentage changes for all
  entries, but for the higher-order Balmer lines
  $\hda,~\hga,~\hdf,~\hgf$ , for which the difference (in~\AA) between
  the value relative to a BHB and that relative to a red HB is given.}
  \begin{tabular}{@{}ccccccccccccccccccc} 
& & & & & & & & & & & & & &  & & & \\ 
${\rm [Z/H]}$ & L$_{U}$ & L$_B$ & L$_V$ & L$_R$ & L$_I$ &
  L$_J$ & L$_H$ & L$_K$ & D4000 & $\rm CaT^{*}$ & $\rm Mg_2$ & Mgb & $<Fe>$ &
  $\hb$ & $\hda$ & $\hga$ & $\hdf$ & $\hgf$ \\ 
& & & & & & & & & & & & & & & & & \\ 
$-0.33$ & 60 & 38 & -8 & -30 & -41 & -52 & -56 & -55 & -21 & -8 & -6 & -17 & -19 & 77 & 5 & 2.9 & 6.4 & 3.35 \\ 
0.00 & 58 & 38 & 5 & -8 & -15 & -24 & -26 & -26 & -25 & -3 & -7 & -15 & -13 & 61 & 4.8 & 2.4 & 5.7 & 2.9 \\ 
0.35 & 70 & 47 & 12 & -4 & -12 & -26 & -31 & -32 & -25 & -2 & -12 & -18 & -14 & 73 & 5.7 & 2.6 & 6.3 & 3.1 \\ 
\end{tabular}
\end{table*}

Table~\ref{BHBtab} provides the effects of high $Z$ BHB on various SSP
output\footnote{In order to evaluate the effect on several other SSP
output the readers can compare the models with BHB and RHB that are
available at the model web page.}. The impact is generally expressed
as a fractional change, i.e. is the percentage variation as referred
to the red HB SSP with its sign. For example, at $\zh=-0.33$ L$_{U}$
increases by 60 per cent , while L$_{K}$ decreases by 55 per cent. For
the higher-order Balmer lines $\hda,~\hga,~\hdf,~\hgf$ (Worthey \&
Ottaviani~1997) whose values cross zero when their strengths decrease,
the difference (in \AA) between the value with BHB and RHB is
given. For example, at the same metallicity, $\hda$ increases by 5
\AA~ when a BHB is present. The obvious consequence of having a blue
HB morphology is to enhance the strength of the spectro-photometric
indicators in the blue. From this effect comes the age-metallicity
degeneracy due to the Horizontal Branch (e.g. Maraston \&
Thomas~2000), that is that an unresolved BHB in an old stellar
population can mimic a substantially younger age. For example, a 10
Gyr SSP with solar metallicity and a BHB would appear as young as 2
Gyr in the Balmer lines. Note that also the metallic indices as well
as the D-4000 and the CaII index change consistently and would yield
an age of approximately 2 Gyr. The effect on index-index diagrams is
shown in Thomas et al.~(2004).

In order to break the age-BHB degeneracy the best strategy appears to
be a combination of Balmer lines (or blue colours) and TP-AGB
sensitive colours like $H-K$. Indeed as one sees from Table 3 the
latter is insensitive to the presence of a BHB while it increases
substantially in presence of a real 2 Gyr old population containing
TP-AGB stars (Figure~20).

The values provided in Table~3 allow the evaluation of BHB effects in
arbitrary proportions on a stellar population, also when other models
from the literature are adopted. In particular, the values in the
Table can be applied to the models with variable element abundance
ratios by Thomas, Maraston \& Bender~(2003, for the Lick indices) and
by Thomas, Maraston \& Korn~(2004, for the higher-order Balmer lines).

\section{Uncertainties on SSPs due to stellar tracks or EPS codes}
\label{uncertainties}
\begin{figure*}
 \psfig{figure=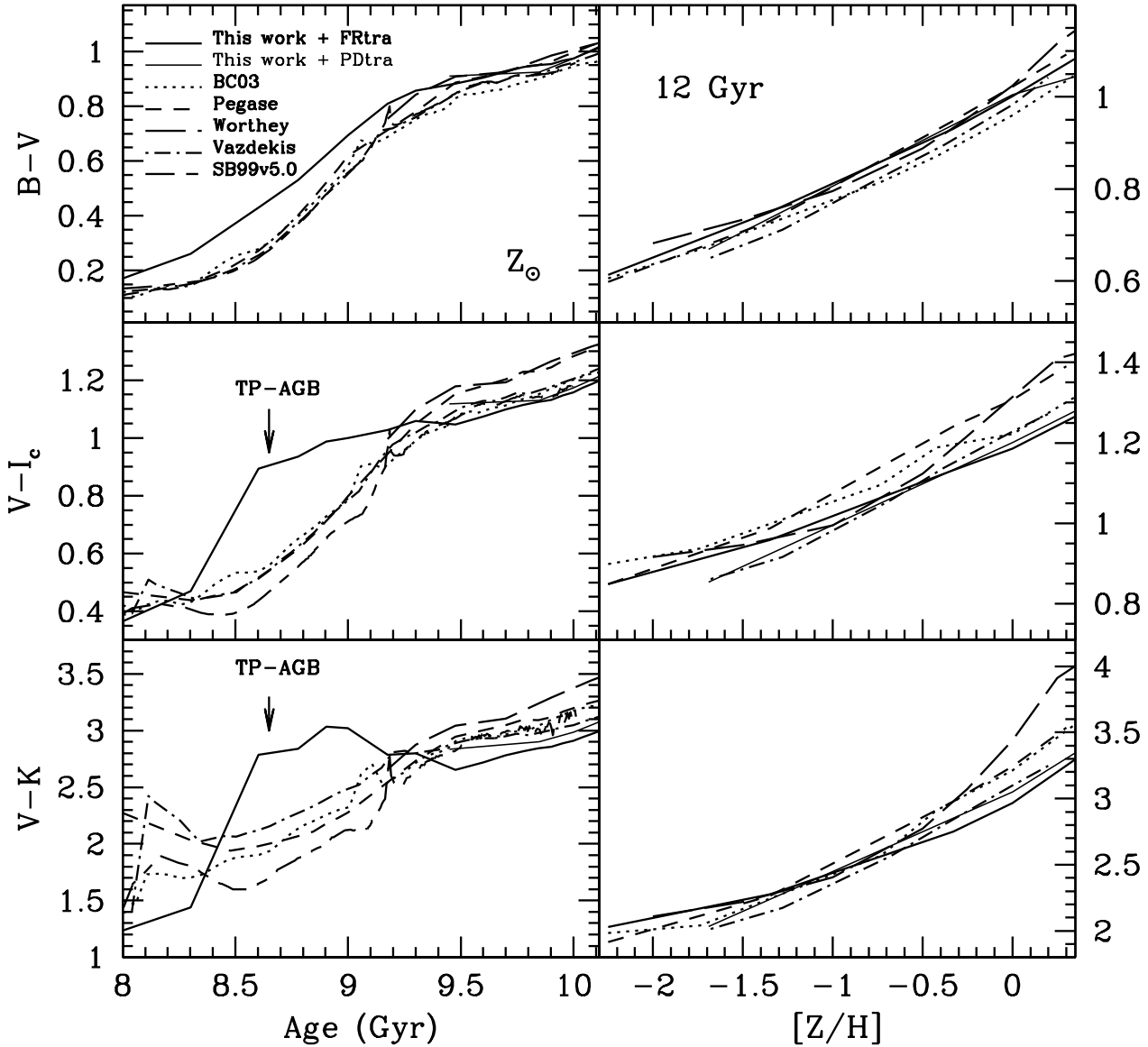,width=\linewidth} \caption{Effects of
  stellar models and EPS codes on colour-age (left-hand panel) and
  colour-metallicity at fixed (12 Gyr) relations of stellar population
  models. The thin solid lines are SSP models computed with our code
  and the Padova stellar evolutionary tracks. The effect of the TP-AGB
  phase (see Sections 3.4.3 and 4 and Figures 18, 19 and 20) is
  highlighted.}
\label{complito}
\end{figure*}
Two types of uncertainties affect the interpretations of the age and
the metallicity of an unresolved stellar population by means of SSPs
models. The first group collects what we can call {\it intrinsic
uncertainties}, in the sense that they belong to the physiology of the
model. These are led by the ``age-metallicity degeneracy''
(e.g. Faber~1972; O'Connell 1980; Worthey 1994; Maraston \&
Thomas~2000), the effect by which a larger metal content produces
similar spectral changes as an older age, and vice versa. To this
group belongs the HB morphology and the chemical abundance
pattern. These uncertainties cannot be removed, however they can be
taken into account by a proper use of the models.

\begin{figure*}
 \psfig{figure=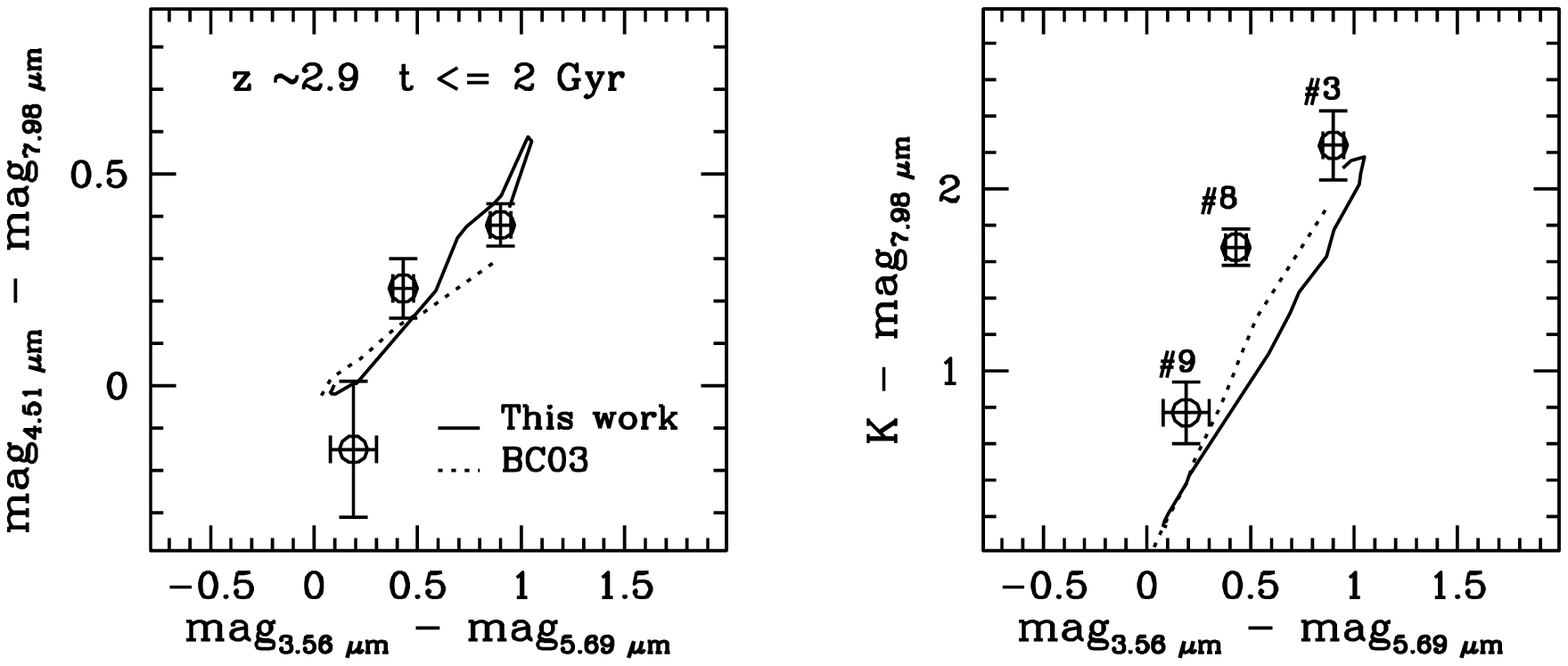,width=\linewidth}
  \caption{Two-colour diagrams in the IRAC colours (left-hand panel)
  and IRAC and $K$ (right-hand panel), of the three highest-$z$
  objects (\# 3,\# 8,\# 9) of the Yan et al.~(2004) sample. Magnitudes
  are in the AB system, effective wavelengths of the IRAC filters are
  indicated. The $y$-axis of the right-hand panel corresponds to the
  rest-frame $V-K$. Dotted and solid lines refer to the models by BC03
  (solar metallicity) and ours for $0.5~\zsun$ plotted until the age
  of 2 Gyr, in order not to exceed the local age of the Universe
  (adopted cosmology $\Omega_{\rm M}=0.3, \Omega_{\rm \lambda}=0.7,
  \rm H_{\rm 0}=73~\rm km \cdot s^{-1} Mpc^{-1}$). }
\label{irac_col}
\end{figure*}
The other class should be called {\it transient uncertainties}, in the
sense that a solution can be found at some point. To this group belong
the discrepancies in the model output that are generated by: 1)
stellar evolutionary tracks; 2) transformations to the observables; 3)
EPS codes. Charlot, Worthey \& Bressan~(1996) investigated the origin
of the discrepancies in their model $B-V$, $V-K$ and $M/L_{\rm
V}$. Their main conclusions are: a) the 0.05 mag discrepancy in $B-V$
colour originates from a known limitation of the theoretical spectra;
b) the 0.25 mag discrepancy in $V-K$ comes from the stellar evolution
prescriptions; c) lifetimes and luminosities of post-MS phases are
problematic. Since different EPS codes were used, their approach did
not allow the exploration of the sole effect of stellar models. This
is what we shall do here. With the same code we vary the stellar
evolutionary input, i.e. the - energetics and temperatures - matrices
while keeping constant the temperature/color/spectra-transformation
matrix. A similar exercise can be found in Fioc \&
Rocca-Volmerange~(1997), where the impact of varying the stellar
tracks from the Padova to the Geneve set is tested. Major differences
were found at very young ages (tenths of Myr), that are connected with
the recipes of mass-loss in massive stars. However both sets of tracks
adopt overshooting, while in the exercise presented here the Frascati
tracks do not include this parameter, thereby allowing to explore its
impact on the spectral evolution. In the following we will focus on
colours and spectra, since the influence of the stellar tracks on the
Lick absorption indices has been discussed by Maraston et al.~(2001b;
2003). We also leave out the investigation of the uncertainties in the
spectral libraries since it is made in BC03.

In Figure~\ref{complito} (left-hand panel) we compare selected
broadband colours with solar metallicity and various ages as computed
with our code and the Frascati and the Padova stellar tracks (solid
thick and thin lines, respectively). 

The $V-K$ is the most affected by the stellar models. At $t \ga 2$ the
value obtained with the Padova tracks is redder by 0.08 mag than that
obtained with the Frascati tracks. This was expected since the largest
difference among the two sets of tracks is the RGB temperature
(Figure~\ref{tergbtracks}). The difference is not dramatic since the
tracks deviate especially towards the tip, where the evolutionary
timescale is faster, therefore the fuel consumption smaller. Similar
values are found at twice solar metallicity. 

The $V-I$ is rather insensitive to the choice of the
tracks\footnote{Note that our SSPs based on the Padova tracks do not
include the TP-AGB phase therefore the differences at $t\lapprox~2~\rm
Gyr$ are not relevant in this context.}, while the $B-V$ relative to
the Padova tracks is bluer at the ages at which the overshooting is
important, in agreement with the higher MS luminosity of the Padova
tracks (Figure~7). At old ages the two model $B-V$ are in perfect
agreement. As for the metallicity scale (right-hand panels), the
largest discrepancies are found when the colors involving near-IR
bands are considered, which again stems from the different RGB
temperatures. For example at low metallicities the Padova tracks are
hotter than the Frascati ones, and the relative colours
bluer. Therefore an observed colour (at given age) is interpreted as
connected to a higher metallicity (e.g. a $V-K$ of 2 requires $\zh
\sim -1.7$ for SSPs based on the Padova tracks, but just $\zh \sim
-2.2$ for those based on the Frascati tracks). The differences are
much smaller in the optical bands.

Also plotted are models by other EPS codes (Worthey~1994, Vazdekis et
al.~1996, PEGASE.2, BC03 and V\'azques \& Leitherer 2005). All these
models except those by Worthey~(1994) adopt the stellar evolutionary
tracks from the Padova database and are computed by means of the
isochrone synthesis tecnique. In spite of such harmony of model
inputs, sizable discrepancies exist between these
models. Interestingly, these type of discrepancies are larger than
those induced by the use of different stellar tracks in our code. For
example, the $V-K$ of the PEGASE.2 models with solar metallicity is
nearly 0.2 mag. redder than that of the Vazdekis models, while BC03
agree with the latter in $V-I$ and with PEGASE.2 in $V-K$. 

Note that in none of these other models the AGB phase transition is
appreciable.

Also the metallicity scale shows quite a discrepancy, especially
approaching large chemical abundances.

The scatter in the inferred ages and metallicities can be estimated
directly from Figure~\ref{complito}. We did not attempt a more
quantitative evaluation since after all metallicities and ages should
not be determined by means of one indicator, rather through a
``grid-like'' approach.

As a final remark, the integrated colours of the Worthey~(1994) models
are typically redder than those of the other models. Charlot, Worthey
\& Bressan~(1996) attribute the effect to a factor 2 more stars on the
upper RGBs of G. Worthey isochrones than on the Padova isochrones used
in the other two EPS. The fuel consumption approach helps to avoid
such uncertainties.
\section{A jump to high-$z$: TP-AGB stars in Spitzer galaxies?}
\label{highztpagb}
As discussed in Section~\ref{energetics} the onset and development of
the TP-AGB phase occupies a narrow age range in the evolutionary path
of stellar populations, being confined to ages $0.3~\lapprox t/{\rm
Gyr} \lapprox 2$. During this short epoch the TP-AGB phase is the
dominant one in a stellar population providing $\sim 40$ per cent of
the bolometric contribution, and up to $\sim 80$ per cent of that in
the $K$-band (Figure~\ref{contrpha}). Therefore the inclusion of the
TP-AGB phase in a stellar population model is essential for a correct
interpretation of galaxies with stellar populations in this age range
as discussed in Section~\ref{compsedTP-AGB}. In particular, the TP-AGB
phase can be used as an age indicator of intermediate-age stellar
populations and its power is to be relatively robust against the
age/metallicity degeneracy. In this direction goes the work by Silva
\& Bothun~(1998), in which the two AGB-sensitive colours J$-$H and
H$-$K were used to constrain the amount of intermediate-age stars in
local, disturbed, field ellipticals. The authors rely on the
comparison with colours of real TP-AGB stars after noticing that the
population models they explored (Worthey 1994 and Bruzual \& Charlot
1993) neither reproduced the near-IR colours of local ellipticals nor
were in agreement one with each other (the latter is confirmed by our
comparison, see Figure~27). Here we add that those models would not be
suitable to the aim since they do not include the calibrated
contribution by the TP-AGB phase (see 4.2.2). The possible use of the
TP-AGB as age indicator for high-$z$ galaxies was firstly suggested by
Renzini~(1992).

At high redshifts when galaxies are dominated by $\sim$ 1 Gyr old
populations as inferred from the ages and element ratios of local
early-type galaxies (Thomas et al.\ 2004), the TP-AGB signature in the
rest-frame near-IR must show up, with e.g. the rest-frame $V-K$ colour
mapping into the observed $K-10$ $\mu$m at $z=3$. This portion of the
spectrum became recently available thanks to the advent of the Spitzer
Space Telescope (SST). Therefore a straightforward application of the
model SEDs presented in this paper is the interpretation of the
high-$z$ Spitzer galaxies as also discussed in Maraston (2004).

Recently Yan et al.~(2004) published SST-IRAC data of galaxies in the
Hubble Ultra Deep Field, with photometric redshifts ranging from 1.9
to 2.9.  In order to explain the observed SEDs in the whole spectral
range from the rest-frame $B$ to $K$, in particular the high fluxes in
both the blue and the near-IR, the authors must combine a dominant
stellar population being at least 2.5 Gyr old, with traces (less than
1 per cent in mass) of a 0.1 Gyr old one. The relatively high age of
the (dominant!) old component generally implies very high formation
redshifts ($z_{\rm f}>>10$), and even exceeds the age of the universe
for the highest redshift objects.

This contrived solution is imposed by the evidence of very high
observed fluxes in both the optical and near-IR rest frames.  As we
saw in Section~\ref{fingerTP-AGB} and Figure~19, high near-IR fluxes
in young stellar populations are the fingerprints of TP-AGB stars. It
is therefore interesting to explore which ages are inferred from the
models of this paper.

Figure~\ref{irac_col} shows two-colour diagrams in the IRAC+$K$
observed frame, for the three highest-$z$ objects of the Yan et
al.~(2004) sample. The $y$-axis in the right-hand panel corresponds to
the rest-frame $V-K$, that is very sensitive to TP-AGB stars
(cf. Figure~20). At the nominal redshift of the objects ($z \sim 2.9
$) the universe is at most 2.3 Gyr old in current cosmologies. The
BC03 models (dotted lines) used by Yan et al.~(2004) do not match the
intrinsic $V-K$ of the reddest object (\#3, right-hand panel) unless
the age becomes 4 Gyr, that exceeds the age of the universe. Our model
(solid line) reaches the observed colours using ages around
$\sim~1.5-2~\rm Gyr$, instead. The inferred stellar mass for object
\#3 is $\sim 2.5~10^{10}~\msun$. The other two objects in
Figure~\ref{irac_col} look younger in this simple diagrams and seem to
be in the pre-TP-AGB phase. However, object \#8 has a rest-frame $V-K$
that is too red (or a rest-frame $I-J$ that is too blue, x-axis) also
for our standard models. A possible explanation could be reddening by
dust, that would affect mostly the rest-frame $V$. Dust reddening was
not found to be a promising solution by Yan et al.~(2004), however
this might be also connected with their modelling lacking
intrinsically red stars in the stellar population. A more detailed
investigation is clearly beyond the scope of this paper and will be
addressed in a subsequent work.

\begin{figure}
 \psfig{figure=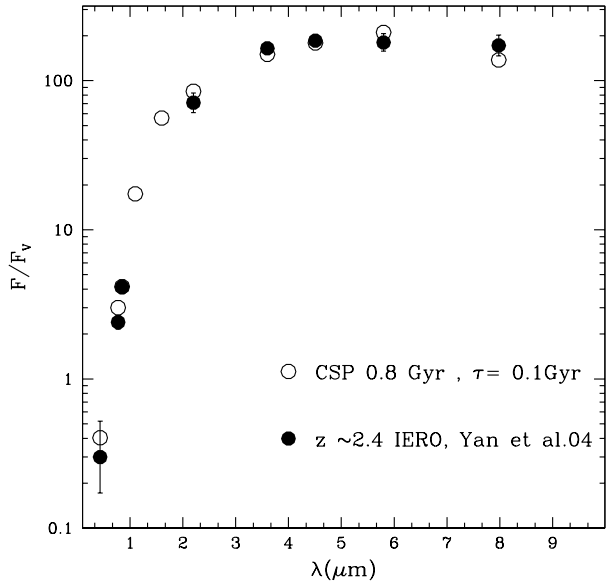,width=\linewidth}
  \caption{Comparison between the observed narrow-band fluxes of
  object \#11 from Yan et al.~(2004) and those of a 0.8 Gyr old
  stellar population model with solar metallicity in which the stars
  are forming with an exponentially-declining star formation rate with
  e-folding time of 0.1 Gyr.}
\label{irac_fit1}
\end{figure}
\begin{figure}
 \psfig{figure=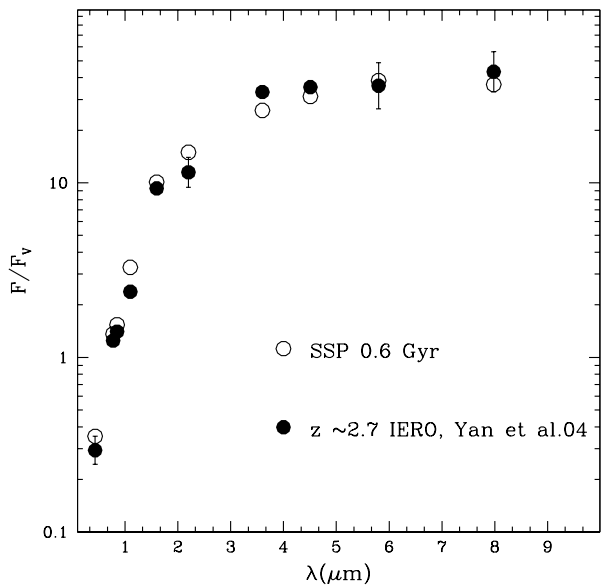,width=\linewidth}
  \caption{Comparison between the observed narrow-band fluxes of
  object \#15 from Yan et al.~(2004) and those of a 0.6 Gyr old
  metal-poor simple stellar population.}
\label{irac_fit2}
\end{figure}
As a further example Figures~\ref{irac_fit1} and \ref{irac_fit2} show
the fits to the whole SEDs for other two objects of the Yan et
al.~(2004) sample. In the first case the best ``by-eye'' solution was
found to be a composite stellar population (with solar metallicity) in
which stars started to form 0.8 Gyr ago, with an
exponentially-declining star formation mode with e-folding time of 0.1
Gyr. In the second case, a single burst (i.e. an SSP) of 0.6 Gyr with
low-metallicity ($\zsun/20$) was instead used. For both objects the
whole spectral energy distributions are fitted reasonably well with
lower ages, corresponding to formation redshifts of the order $z\sim
3$. The derived stellar masses are $1.56 \cdot~10^{10}~\msun$ and
$1.54 \cdot~10^{9}~\msun$, respectively. Interestingly, in the latter
case the mass and the metallicity fits very well the corresponding
relation obtained in the local universe for dwarf early-type galaxies
(Thomas et al.~2003b). This example opens the possibility that at
least part of the very red galaxies could be objects caugth during the
full development of the TP-AGB phase.

The fits shown in Figures~\ref{irac_fit1} and \ref{irac_fit2} are
quite rudimentary and are meant to illustrate the application of the
new models to high-$z$ data. The `best' solution was searched only
``by eye'', without the help of statistical tools, which would go far
beyond the scope of the paper. However the match is encorauging and we
will analyse more objects and in a more precise way in a future work.
\section{Summary}
Evolutionary population synthesis models with various ages,
metallicities, star formation histories, Horizontal Branch
morphologies, are presented. The evolutionary code is based on the
fuel consumption theorem (Renzini \& Buzzoni~1986) for the evaluation
of the energetics of the post Main Sequence evolutionary phases. The
code was introduced in our previous work (Maraston~1998) in which
stellar population models complete in all main stellar evolutionary
phases, but only for solar metallicity were computed. The present
paper provides the extension of the modelling to a wider range of
stellar population parameters. The most important features of the
models are: i) the inclusion of the energetics and the spectra of
TP-AGB stars and their calibrations with observations; ii) the
allowance for various Horizontal Branch morphologies, in particular
blue morphologies at high metallicities. We will come back to both
points below.

In the first part of the paper we perform a comprehensive analysis of
the model ingredients, focusing in particular on: i) the metallicity
effects on energetics and temperatures; ii) the impact of different
stellar evolutionary tracks. The most relevant results can be
summarized as follows.

The energetics of stellar populations are higher at lower metallicity
because of the greater hydrogen abundance. The most affected stellar
phases are the main sequence and the helium burning phase. The
differences become smaller at ages later than the RGB phase-transition
($t\gapprox~1~\rm Gyr$) when the RGB phase becomes the most important
contributor to the total bolometric. This is due to the RGB fuel
consumption being not affected dramatically by the chemical abundance,
a result consistent with previous findings (Sweigart, Greggio \&
Renzini~1989). 

The impact of stellar evolutionary tracks is evaluated by comparing
energetics and temperatures of two different stellar models, namely
the set of Frascati tracks by Cassisi and collaborators, that are
classical tracks without overshooting, with those by the Padova
school, that include the effects of overhsooting. We find that the
mass- luminosity relations for MS stars are rather similar among the
explored tracks, which is explained by the modest amount of
overshooting included in recent computations. Instead, major
differences are found in post-MS phases, which involve both the
energetics and the temperatures of the Red Giant Branch. In the Padova
tracks the onset and development of the RGB is delayed with respect to
the Frascati tracks, an effect of overshooting. For example, a
$\sim 0.8~\rm Gyr$ stellar population model based on the Padova tracks
has $\sim 30$ per cent less light than one based on the Frascati
tracks, a result influencing the integrated output, especially the
$M/L$. It should be noticed that a recent comparison of the onset age
of the RGB with what is observed in MC GCs favours the earlier RGB
development of the Frascati tracks (Ferraro et al.~2004).

The temperatures of the RGB, especially towards the tip are cooler in
the Padova tracks, at solar metallicity and above, a feature also
found in the Yale tracks of Yi et al.~(2003). The RGB temperature is
the ingredient having the largest influence on the SSP output. For
example, the integrated $V-K$ colours of old, metal-rich, stellar
populations change by $\Delta V-K\sim 0.08~\rm mag$. The RGB
temperatures are connected to the mixing-length and the underlying
stellar models. The calibration of these effects is difficult since
the comparisons with data require the adoption of temperature-colors
transformations, that are uncertain as well. In spite of this, it
would be highly valuable to understand the origin of such
discrepancies at least theoretically. 

Relevant to evolutionary population synthesis models is the following
result. The differences induced by the use of various stellar tracks
in our code are substantially smaller than the scatter among stellar
population models in the literature that adopt the same stellar
tracks.

\medskip
In the second part of the paper we describe in detail the recipes for
the stellar phases that are affected by mass-loss and therefore not
described sufficiently by stellar evolutionary tracks. These are the
Horizontal Branch (HB) in old stellar populations and the
Thermally-Pulsing Asymptotic Giant Branch (TP-AGB) in stellar
populations with ages between 0.2 and 2 Gyr. For the HB we apply
mass-loss to the RGB tracks such that we obtain a good match to
stellar population spectrophotometric indicators that are sensitive to
the HB morphology (e.g. the Balmer lines). We therefore calibrate the
amount of mass-loss with MW GCs of known HB morphology. In order to
account for the observed scatter in the HB properties at a fixed
metallicity, we provide the models with a couple of choices for the HB
morphology. In particular, we also compute blue HB at high $Z$ and
show how much blue indicators like e.g. the Balmer lines and the
$U,B,V$ magnitudes are affected. The effects of BHBs at high $Z$ are
quantified in a form that can be easily used also when other
evolutionary population synthesis are adopted. 

A result interesting for extra-galactic GC studies is that the D-4000
\AA~break, unlike the Balmer lines is insensitive not only to age but
also to the Horizontal Branch morphology at low metallicity ($\zh
\lapprox -1$). Therefore this colour can be used as metallicity
indicator in the metal-poor regime.

The energetic of the TP-AGB phase was calibrated with Magellanic
Clouds GCs data in Maraston (1998). It was shown that this phase is
the dominant one in stellar populations with ages between 0.3 and 2
Gyr, providing 40 per cent of the bolometric and up to 80 per cent of
the luminosity in the near-IR. Here we make a step forward by
including the spectra of TP-AGB stars in the synthetic spectral energy
distributions (SEDs) of the stellar population models, using available
empirical spectra of TP-AGB C-rich and O-rich stars. We compare the
resulting model SEDs with the observed SEDs of a sample of MC GCs with
data from $U$ to $K$. The models provide a very good fit to the data
in the whole age range relevant to the TP-AGB, a success that is
entirely due to the proper inclusion of the phase in the models. We
further compute the integrated spectral indices $\rm C_{2}, H_{2}O,
CO$ in the near-IR ($1.7~\lapprox \lambda/\rm \mu m \lapprox 2.4$),
that are effective tracers of C and O stars. We find that the
combination of $\rm C_{2}$ and $\rm H_{2}O$ can be used to determine
the metallicity of $\sim 1~\rm Gyr$ stellar populations, independently
of the star formation history.

The model SEDs presented here can be applied to the analysis of
high-$z$ galaxies extending the concept of TP-AGB as age indicator for
$\sim~1~{\rm Gyr}$ stellar populations to primeval objects in the
early universe. As an illustrative application we re-analyze some
high-$z$ ($2.4 \lapprox z\lapprox 2.9$) galaxies recently observed
with the Spitzer Space Telescope (Yan et al.~2004). The distinctive
features of these objects is to have strong fluxes in both the
rest-frames blue and near-IR relative to the flux in the
$V$-band. This is what characterizes TP-AGB dominated stellar
populations. We show that the data can be explained by stellar
populations with ages in the range 0.6 to 1.5 Gyr. These ages are
comfortable given the age of the universe at those redshifts
($\sim~2~\rm Gyr$) and imply formation redshifts between 3 and 6.

\medskip
Models are available at www-astro.physics.ox.ac.uk/$\sim\,$maraston.

\section*{Acknowledgments}
In the list of acknowledgments the pole position is clearly gained by
Santino Cassisi, for his constant availability to satisfy the most
exotic requests in terms of evolutionary computations. Many thanks go
to Laura Greggio, Alvio Renzini, and Daniel Thomas for the careful
reading of the manuscript, and their very constructive comments. I
would also like to thank Paul Goudfrooij for providing the globular
cluster data before publication. Furthermore I am grateful to Don
Vandenberg for very useful comments on temperature/colors
transformations and for having provided some of his models in advance
of publication.

\end{document}